\documentclass[twocolumn]{aastex63}

\newcommand{\beq}{\begin{equation}}
\newcommand{\eeq}{\end{equation}}
\usepackage{amsmath}
\usepackage{longtable}

\shorttitle{Discovery of dust enrichment and grain growth in a smooth DG Tau disk }
\shortauthors{Ohashi et al.}
\graphicspath{{./}{figures/}}

\begin{document}

\title{Dust enrichment and grain growth in a smooth disk around the DG Tau protostar revealed by ALMA triple bands frequency observations}

\author[0000-0002-9661-7958]{Satoshi Ohashi}
\altaffiliation{NAOJ Fellow}
\affil{National Astronomical Observatory of Japan, 2-21-1 Osawa, Mitaka, Tokyo 181-8588, Japan}
\affil{RIKEN Cluster for Pioneering Research, 2-1 Hirosawa, Wako-shi, Saitama 351-0198, Japan}
\email{satoshi.ohashi@nao.ac.jp}

\author[0000-0002-3001-0897]{Munetake Momose}
\affiliation{College of Science, Ibaraki University, 2-1-1 Bunkyo, Mito, Ibaraki 310-8512, Japan}

\author[0000-0003-4562-4119]{Akimasa Kataoka}
\affiliation{National Astronomical Observatory of Japan, 2-21-1 Osawa, Mitaka, Tokyo 181-8588, Japan}

\author[0000-0002-9221-2910]{Aya E Higuchi}
\affil{Division of Science, School of Science and Engineering, Tokyo Denki University, Ishizaka, Hatoyama-machi, Hiki-gun, Saitama 350-0394, Japan}

\author[0000-0002-6034-2892]{Takashi Tsukagoshi}
\affiliation{Faculty of Engineering, Ashikaga University, Ohmae-cho 268-1, Ashikaga, Tochigi 326-8558, Japan}

\author[0000-0003-4902-222X]{Takahiro Ueda}
\affiliation{Max-Planck Institute for Astronomy, K\"{o}nigstuhl 17, D-69117 Heidelberg, Germany}

\author[0000-0003-1514-3074]{Claudio Codella}
\affil{INAF, Osservatorio Astrofisico di Arcetri, Largo E. Fermi 5, I-50125, Firenze, Italy}
\affil{Univ. Grenoble Alpes, CNRS, IPAG, 38000 Grenoble, France}

\author{Linda Podio}
\affiliation{INAF, Osservatorio Astrofisico di Arcetri, Largo E. Fermi 5, I-50125, Firenze, Italy}

\author[0000-0002-7538-581X]{Tomoyuki Hanawa}
\affil{Center for Frontier Science, Chiba University, 1-33 Yayoi-cho, Inage-ku, Chiba, Chiba 263-8522, Japan}

\author[0000-0002-3297-4497]{Nami Sakai}
\affil{RIKEN Cluster for Pioneering Research, 2-1 Hirosawa, Wako-shi, Saitama 351-0198, Japan}

\author[0000-0001-8808-2132]{Hiroshi Kobayashi}
\affil{Department of Physics, Graduate School of Science, Nagoya University, Furo-cho, Chikusa-ku, Nagoya 464-8602, Japan}

\author[0000-0002-1886-0880]{Satoshi Okuzumi}
\affil{Department of Earth and Planetary Sciences, Tokyo Institute of Technology, Meguro-ku, Tokyo 152-8551, Japan}

\author[0000-0001-9659-658X]{Hidekazu Tanaka}
\affil{Astronomical Institute, Tohoku University, 6-3 Aramaki, Aoba-ku, Sendai 980-8578, Japan}




\begin{abstract}

Characterizing the physical properties of dust grains in a protoplanetary disk is critical to comprehending the planet formation process. Our study presents ALMA high-resolution observations of the young protoplanetary disk around DG Tau at a 1.3 mm dust continuum. The observations, with a spatial resolution of $\approx0\farcs04$, or $\approx5$ au, revealed a geometrically thin and smooth disk without substantial substructures, suggesting that the disk retains the initial conditions of the planet formation. To further analyze the distributions of dust surface density, temperature, and grain size, we conducted a multi-band analysis with several dust models, incorporating ALMA archival data of the 0.87 mm and 3.1 mm dust polarization. The results showed that the Toomre {\it Q} parameter is $\lesssim2$ at a 20 au radius, assuming a dust-to-gas mass ratio of 0.01. This implies that a higher dust-to-gas mass ratio is necessary to stabilize the disk. The grain sizes depend on the dust models, and for the DSHARP compact dust, they were found to be smaller than $\sim400$ $\mu$m in the inner region ($r\lesssim20$ au), while exceeding larger than 3 mm in the outer part. Radiative transfer calculations show that the dust scale height is lower than at least one-third of the gas scale height. These distributions of dust enrichment, grain sizes, and weak turbulence strength may have significant implications for the formation of planetesimals through mechanisms such as streaming instability.  We also discuss the CO snowline effect and collisional fragmentation in dust coagulation for the origin of the dust size distribution.

\end{abstract}



\section{Introduction} \label{sec:intro}

Protoplanetary disks consist of gas and dust around a young star and are considered to be the birthplace of planets. To understand the process of planet formation, it is crucial to constrain the physical conditions of dust grains in disks since dust grains are the building blocks of planets.
In the standard scenario for planet formation, dust grains grow by coagulation from micron-sized particles present in the molecular cloud cores to centimeters and even larger bodies on their way to becoming planetesimals \citep[e.g., ][and references therein]{tes14}. 
However, millimeter-sized dust grains formed in a few tens au radius quickly fall toward the central star due to the gas pressure \citep{ada76}.
Several mechanisms have been proposed to overcome this radial drift problem, such as a gas pressure bump to accumulate the dust grains \citep{pin12} or rapid coagulation of highly porous aggregates \citep{oku12,kob21}.

Recent observations with the Atacama Large Millimeter/submillimeter Array (ALMA) have revealed that axisymmetric gap and ring structures are common in the dust continuum emission of protoplanetary disks \citep[e.g.,][]{alma15,and18}.
These substructures may play an important role in planet formation because dust grains would accumulate in the substructure regions, such as the positions of the rings, where grain growth occurs without the radial drift of large dust grains, although the mechanism of substructure formation is still under discussion. The presence of planets has also been suggested in the gap regions as detections of a point source emission in near infrared observations \citep[e.g.,][]{kep18} and velocity perturbations of rotation motion in molecular line observations \citep[e.g.,][]{tea18,pin19}.

To elucidate the formation mechanism of the ring-gap structure and to understand the onset of planet formation in terms of grain growth, a structureless smooth disk is an ideal target to investigate the initial conditions of planet formation.
 Even if planets are the origin of the substructure formation, a smooth disk will be important for the initial conditions of planet formation because planets would not have formed in the smooth disk yet.

A young protoplanetary disk around DG Tau, also classified as a protostellar disk, is one of the most promising targets for studying the early stages of the dust disk. DG Tau is classified as a  Class I$-$II protostar with an age of $\sim1$ Myr, a stellar mass of $\sim0.7$ $M_\odot$, and a mass accretion rate of $\sim4.6\times10^{-8}-7.4\times10^{-7}$ $M_\odot$ yr$^{-1}$ \citep{whi01,whi04,ake05,fur06,alc21}.
The results of these early stages of star formation in DG Tau are similar to those of HL Tau, which shows multiple ring and gap structures \citep{alma15}.
The distance is estimated to be $\approx125$ pc \citep{gaia20}.
The disk around DG Tau has been imaged with interferometric observations using the IRAM Plateau de Bure Interferometer \citep{dut96}, the Nobeyama Millimeter Array \citep{kit96}, and the Combined Array for Research in Millimeter-wave Astronomy \citep{ise10}.
These observations show that the resolved-out emission is not found between different uv coverages, suggesting that all of the dust continuum emission comes from the disk rather than the envelope \citep{loo00}. \citet{kit96} also suggested that the enveloping materials may be dissipating.

ALMA observations have imaged the detailed structures of the DG Tau disk in both dust continuum emission including polarization mode \citep{bac18,har19} and molecular line emission \citep{pod20,gar21,gar22}.
The 0.87 mm dust polarization emission can be explained by self-scattering of the dust thermal emission proposed by \citet{kat15} and \citet{yan16} because the polarization vectors are parallel to the disk minor axis with a polarization fraction of $\approx0.4$\% \citep{bac18}.
In contrast, the 3.1 mm dust polarization emission is suggested to be consistent with an expectation of thermal emission from radiatively aligned non-spherical dust grains \citep{taz17} because the polarization vectors are azimuthal with a polarization fraction of $\approx0.5-6$\% \citep{har19}.
These different dust polarization morphologies allow us to constrain the grain size to $\sim100$ $\mu$m.
Several molecular lines have been detected, including the H$_2$CO, CS, and CN molecules, and have shown ring-like structures in all lines with a hole in the central region \citep{pod20}. The hole in the molecular lines may be caused by saturation due to a high optical depth in the dust continuum emission.

\begin{table*}[!ht]
\centering
\vspace{0mm}
\caption{Observing setup}
\begin{tabular}{lcccccl}
\hline
\hline
Band & frequency & resolution & polarization & r.m.s & ALMA ID & Reference \\
        &           (GHz) & & & ($\mu$Jy beam$^{-1}$) &\\
\hline 
 7 & 344 & $0\farcs262\times0\farcs147$ & {\it I} & 210  & 2015.1.00840.S & \cite{bac18}\\
             &             &                                    & {\it Q}    & 55 & &  \\
             &             &                                    &   {\it U}   & 62 & & \\
 6 & 230 & $0\farcs04\times0\farcs032$ & {\it I} & 11  & 2015.1.01268.S & This study\\
 3 & 98 & $0\farcs246\times0\farcs218$ & {\it I} &36  & 2017.1.00470.S & \cite{har19}\\
             &             &                                & {\it Q}                 & 19 & & \\
             &             &                                &  {\it  U}                 & 19 & &  \\
\hline
\end{tabular}\\
 \label{tbl:obs}
\end{table*}

Despite a similar evolutionary stage as the HL Tau disk (age $\sim1$ Myr), the DG Tau disk has not shown any significant ring or gap structure in the continuum emission at a resolution of 0\farcs15 (corresponding to 20 au) \citep{ise10,bac18}.
It should be noted that \citet{pod20} identified a ring-like structure at a radius of 30 au from the protostar in the 0.87 mm continuum emission using an unsharp masking technique that artificially increases the contrast of the image.
 Our new high spatial resolution observations can be used to determine the exact morphology of the disk structure.

In this paper, we present ALMA long-baseline observations in 1.3 mm dust continuum emission from the DG Tau disk to investigate the detailed structure of the disk.
In addition, we study the spectral energy distribution (SED) by incorporating the ALMA data obtained by \citet{bac18} and \citet{har19} to characterize the physical properties of the dust grains.
The details of the observations and the data reduction and imaging processes are described in Section \ref{sec:obs}.
The results of the observations are described in Section \ref{sec:res}.
We model the dust disk with a parametric fit to the observed images in Section \ref{sec:analysis}. Discussion and conclusions are given in Sections \ref{sec:dis} and \ref{sec:sum}, respectively.

\section{Observations} \label{sec:obs}

We describe the 1.3 mm dust continuum observations in Section \ref{obs:1.3} and the ALMA archival data of the 0.87 mm and 3.1 mm dust polarization in Section \ref{obs:pol}.
A summary of these observations is given in Table \ref{tbl:obs}.

\subsection{The 1.3 mm dust continuum data}\label{obs:1.3}

DG Tau was observed with ALMA long-baseline configuration on 22, 30, and 31 October and on 1 November 2015 (2015.1.01268.S, PI: A.Higuchi). The array consisted of 31 antennas on October 22 and 39 antennas on the other three days, providing baselines from 38.6 m to 16.1 km. The correlator processed dual polarizations in four spectral windows centered at 221, 223, 237, and 239 GHz. Each window has a bandwidth of 1.875 GHz, resulting in a total bandwidth of 7.5 GHz with a mean frequency of 230 GHz (1.3 mm wavelength). 
The observations were cycled between the target and J0426+2327 with a cadence of 1 minute. J0510+1800 and J0238+1636 were also briefly observed as calibrators. The total observing time on the target was $\approx4$ hours.

The visibility data were reduced and calibrated using the Common Astronomical Software Application (CASA) package, version 4.5.0 \citep{casa22}. 
After applying phase corrections from water vapor radiometer measurements, the data of problematic antennas and spectral channels severely contaminated by strong telluric lines were flagged. 
The bandpass response of each spectral window was calibrated by the data of J0510+1800. 
The amplitude scale was determined by J0510+1800 and J0238+1636, and the complex gain response of the system was calibrated by the frequent observations of J0426+2327.

The continuum image at 230 GHz was generated by Fourier transforming the calibrated visibilities, deconvolution with the CLEAN algorithm, and finally convolving with a synthesized beam. To improve image fidelity, we performed an iterative phase-only self-calibration using the initial CLEAN image as the first model image. The interval time to solve the complex gain was reduced from inf, 600s, and finally to 300s.
Briggs weighting with a robust parameter of 0.5 was applied throughout the above processes, resulting in the synthesized beam of $0\farcs04\times0\farcs032$ (FWHM) with a position angle of $0.51^\circ$.
After correcting for the primary beam attenuation, the final image has an rms noise level of about 11 $\mu$Jy beam$^{-1}$.

\subsection{The 0.87 mm and 3.1 mm dust continuum data}\label{obs:pol}

We use the ALMA archive data for 0.87 mm (343 GHz at Band 7) and 3.1 mm (97.6 GHz at Band 3) dust continuum emission from the DG Tau disk.
The detailed observations and data reduction for ALMA Bands 7 and 3 are described in \citet{bac18} and \citet{har19}, respectively.
The visibility data, in this study, were reduced and calibrated by the CASA pipeline scripts.
The polarization calibration followed the standard procedure described in \citet{nag16} and in the ALMA polarization casa guide.

The Stokes {\it I, Q, U} images at 343 GHz and 97.6 GHz were generated with the CLEAN algorithm and finally convolved with synthesized beams. To improve the image fidelity, iterative phase-only self-calibrations were performed in the same manner as for the 230 GHz data.
Briggs weighting with a robust parameter of 2 was applied to the 343 GHz data, and Briggs weighting with a robust parameter of 0.5 was applied to the 97.6 GHz data.
The resulting synthesized beams of the 343 GHz data and the 97.6 GHz data are $0\farcs262\times0\farcs147$ (FWHM) with a position angle of $-34.2^\circ$ and $0\farcs246\times0\farcs218$ (FWHM) with a position angle of $-7.42^\circ$, respectively.
The final Stokes {\it I, Q, U} images of the 343 GHz data have rms nose levels of about 210 $\mu$Jy beam$^{-1}$, 55 $\mu$Jy beam$^{-1}$, and 62 $\mu$Jy beam$^{-1}$, respectively.
The final Stokes {\it I, Q, U} images of the 97.6 GHz data have rms nose levels of about 36 $\mu$Jy beam$^{-1}$, 19 $\mu$Jy beam$^{-1}$, and 19 $\mu$Jy beam$^{-1}$, respectively.

\section{Results} \label{sec:res}

\begin{figure*}[htbp]
\begin{center}
\includegraphics[width=18.cm,bb=0 0 4000 2165]{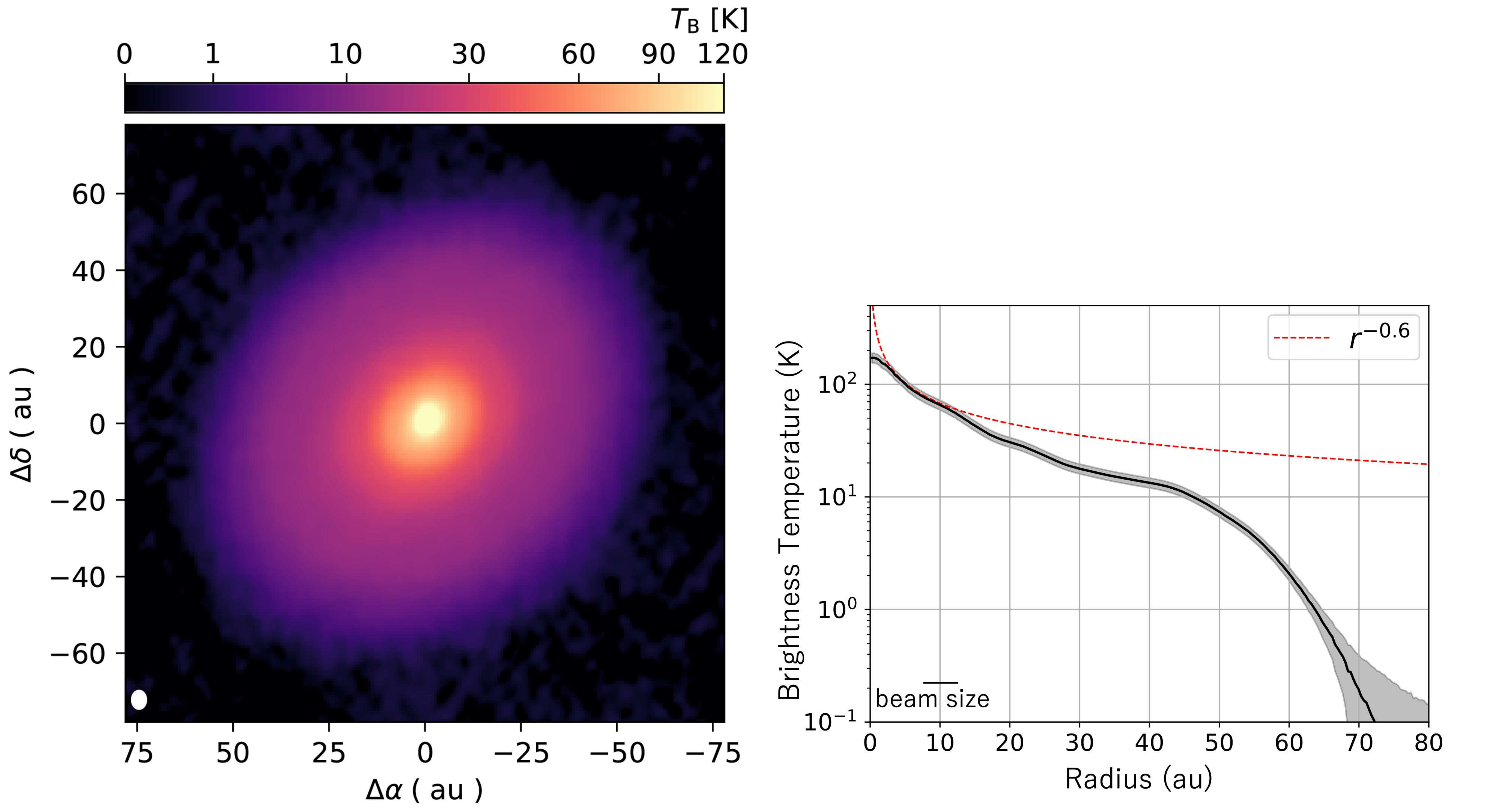}
\end{center}
\caption{(Left) ALMA 1.3 mm dust continuum image. The intensity is converted to the brightness temperature by assuming the Rayleigh-Jeans approximation. The ellipse at the bottom left corner shows the synthesized beam ($0\farcs04\times0\farcs032$).
(Right) A deprojected radial profile of the image. The red dashed lines show a power-law profile of $T_{\rm B}\propto r^{-0.6}$, which is discussed in Section \ref{sec:model1}.
The shadowed region corresponds to the $1\sigma$ error including the rms noise level and 10\% flux uncertainty.
}
\label{fig1}
\end{figure*}

\begin{table*}[!ht]
\centering
\vspace{0mm}
\caption{DG Tau observing parameters}
\begin{tabular}{lcccc}
\hline
\hline
 & peak intensity&   peak brightness temperature$^{a}$   & total flux & polarization fraction  \\
 &  (mJy beam$^{-1}$) & (K)  &  (mJy) & (\%) \\
\hline 
Band 7 &  $197.8\pm0.2$    & $53.3\pm0.1$ & $890\pm10$ & $\lesssim 1$\\
Band 6 & $9.89\pm0.01$   & $178.7\pm0.2$ & $359\pm2$ & \\
Band 3 &$21.22\pm0.03$ & $51.0\pm0.1$   & $48.3\pm0.1$ & $\lesssim 6$ \\
\hline
\end{tabular}\\
 a: The brightness temperature is derived by the Rayleigh-Jeans approximation.
\label{tbl:flux}
\end{table*}

\subsection{ALMA 1.3 mm dust continuum image}

The left panel of Figure \ref{fig1} shows a 1.3 mm dust continuum image of the ALMA long-baseline observations.
The intensity is converted to the brightness temperature using the Rayleigh-Jeans approximation.
The total flux density is estimated to be $359\pm2$ mJy by integrating the emission within the disk region on the image, which is consistent with the previous CARMA observations within the error \citep{ise10}.
The peak position is (R.A. Dec) $=$ (4h27m4.6991s, $26^\circ6'15\farcs738$) with an intensity of $9.89\pm0.01$ mJy beam$^{-1}$ corresponding to $178.7\pm0.2$ K in the brightness temperature ( also shown in Table \ref{tbl:flux}).
The 2D Gaussian fit of the disk image gives a position angle of $135^\circ\pm1^\circ$ and an inclination of $37.3^\circ\pm0.2^\circ$, respectively, consistent with the previous 0.87 mm continuum observations \citep{bac18}.
The disk position angle is almost perpendicular to the jet direction, which consists of a red-shifted velocity component in the northeast of the disk and a blue-shifted velocity component in the southwest \citep{eis98}. The velocity components of the jet indicate that the near and far sides of the disk geometry correspond to the northeast and southwest, respectively.
The dust continuum emission extends to $\approx80$ au from the protostar, which is relatively compact (e.g., $r\approx200$ au in HL Tau) although more compact disks ($r\lesssim50$ au) are common in the low-mass star-forming region \citep{lon19}.

The disk structure of the 1.3 mm continuum emission is well resolved.
 The disk shows the symmetric and smooth structure; we find no significant substructures such as rings or spirals, which are commonly found in a wide range of disk evolution stages from protostellar to protoplanetary disks \citep[e.g.,][]{alma15,and18}.
The smooth morphology is also found by the recent observations from the ALMA Large Program, ``Early Planet Formation in Embedded Disks (eDisk)'', which found relatively few distinct substructures in disks around Class 0/I protostars \citep{oha23}.
Therefore, the DG Tau disk may remain in the initial state before the substructure formation.

The radial profile of the brightness temperature is plotted in the right panel of Figure \ref{fig1}.
The brightness temperature is averaged over a full azimuthal angle.
The error, shown as a shadow, includes the rms noise level and the 10\% flux uncertainty.
The radial distribution shows the smoothed disk structure without significant enhancements or dips as found in the image.
 In addition, the brightness temperature profile shows that the slope of the radial intensity seems to change from the power-law-like distribution to the exponential-like cut-off around $r\approx 45$ au, and this transition position is consistent with the ring-like structure identified by \citet{pod20}.
Therefore, this slope transition was probably observed as the ring-like structure by the unsharp masking technique.
The radial profile of the 1.3 mm emission in DG Tau is not similar to the substructures commonly found in other disks because the intensity gradient decreases monotonically with radius and shows the smooth profile.
The origin of the slope transition is discussed in Section \ref{sec:origin}.
t should be also noted that a small wiggling structure can be found around $r\approx10-30$ au in this profile.
This wiggling may be caused by grain size variations or temperature variations as discussed in Section \ref{sec:model1} and \ref{sec:model2}.

\begin{figure}[htbp]
\begin{center}
\includegraphics[width=8.cm,bb=0 0 1850 1360]{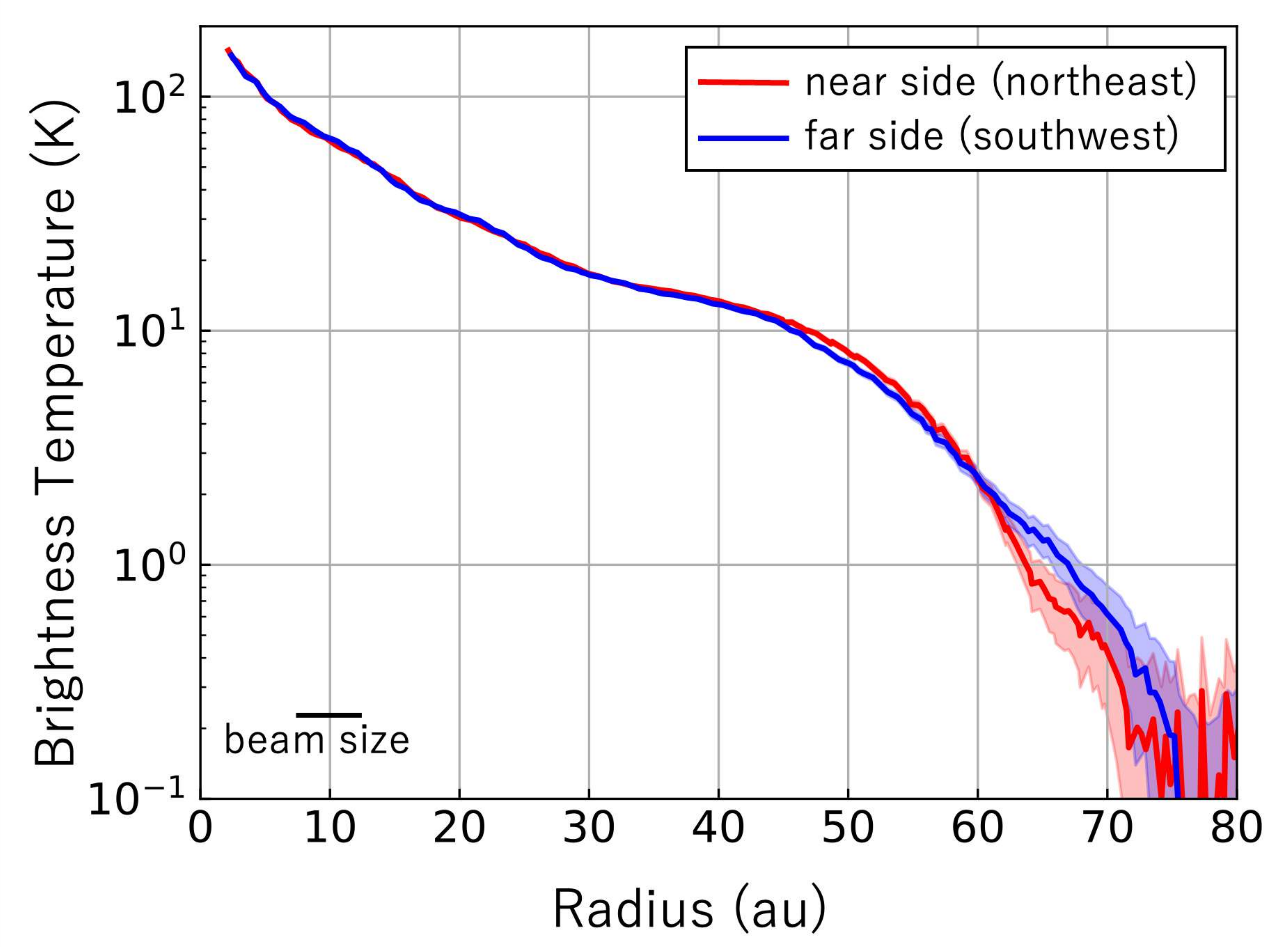}
\end{center}
\caption{Deprojected radial profiles of the ALMA 1.3 mm dust continuum image along the near (red line) and far (blue line) sides.
The position angles of the near and far sides are set to be $40.7^\circ-50.7^\circ$ and $220.7^\circ-230.7^\circ$, respectively.
The shadowed region corresponds to the $1\sigma$ error only including the rms noise level.
}
\label{near_far}
\end{figure}

The finest image also allows us to investigate an asymmetry along the minor axis of the inclined disk between the near and far sides, which will give an implication of the disk flaring effect \citep[e.g.,][]{pin16,vil20}.
Figure \ref{near_far} shows the radial profiles along the far (a position angle of $40.7^\circ-50.7^\circ$) and near (a position angle of $220.7^\circ-230.7^\circ$) sides of the disk minor axis.
The error includes only the rms noise level, since the 10\% flux uncertainty does not affect the relative intensity variations.
Figure \ref{near_far} shows no asymmetry within the 45 au radius, suggesting that dust grains are well settled onto the disk midplane or that the 1.3 mm dust continuum emission is optically thin.
We discuss the dust scale height in more detail in Section \ref{sec:scale_height}.

\subsection{ALMA 0.87 and 3.1 mm dust continuum and polarization images}

\begin{figure*}[htbp]
\begin{center}
\includegraphics[width=17.cm,bb=0 0 3102 1962]{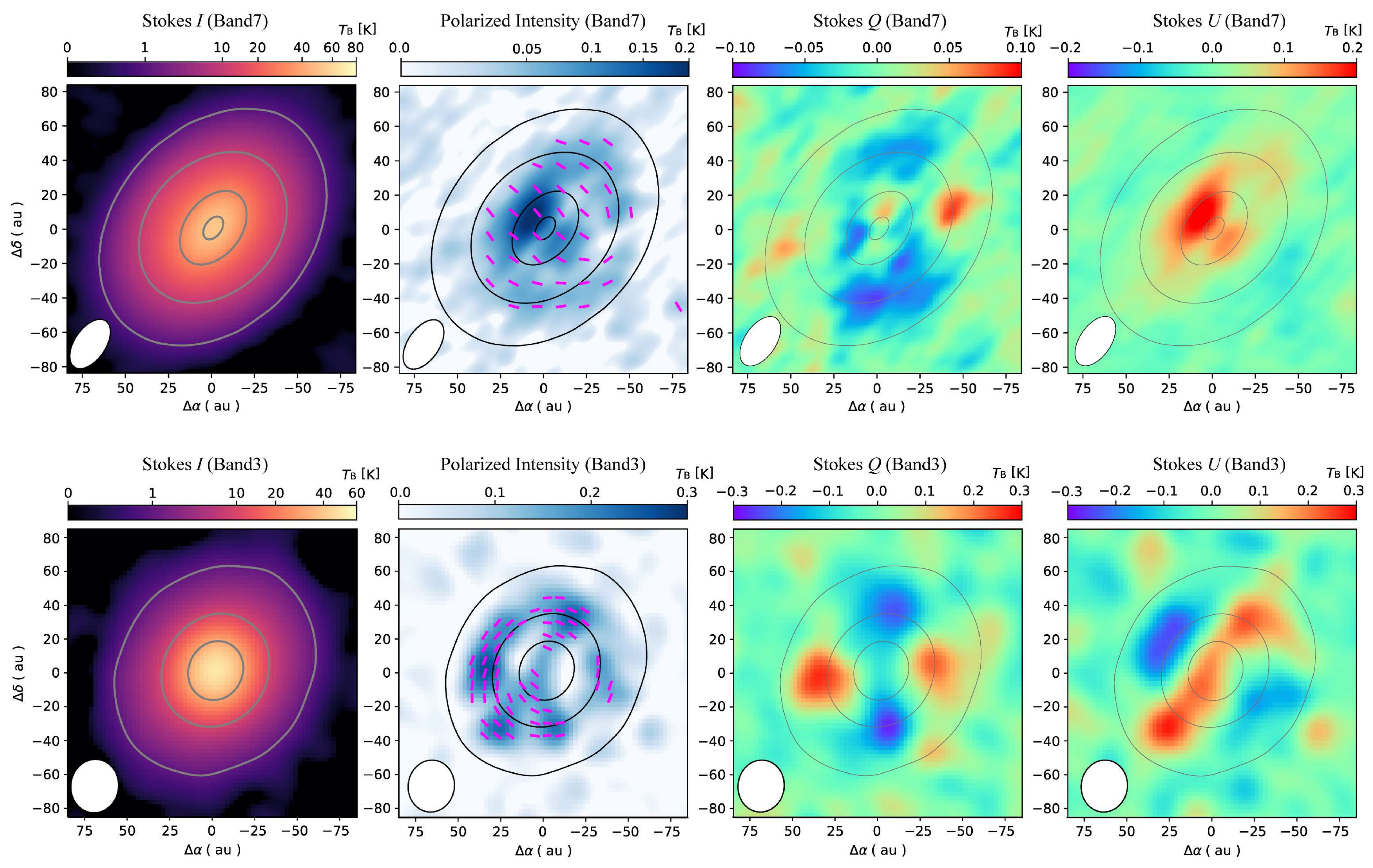}
\end{center}
\caption{ALMA images of the Stokes {\it I}, PI, Stokes {\it Q}, and Stokes {\it U} emission at wavelengths of 0.87 mm (Top panel) and 3.1 mm (Bottom panel). The contours represent the Stokes {\it I} emission of [1, 10, 30, 50] K for 0.87 mm and [1,10,30] K for 3.1 mm.
The bottom left corner of each panel shows the beam sizes.
}
\label{polari_image}
\end{figure*}

The results of the ALMA 0.87 mm and 3.1 mm dust polarization data were already shown by \citet{bac18} and \cite{har19}, respectively.
Here, we briefly show both data and compare the polarization patterns between these wavelengths.
 The peak intensity, total flux, and polarization fraction are compared in Table \ref{tbl:flux}.
Figure \ref{polari_image} shows images of the dust continuum emission (Stokes {\it I}), the linear polarized intensity (PI), and the Stokes parameters {\it QU} for 0.87 mm (top) and 3.1 mm (bottom).
The PI value has a positive bias because it is always a positive quantity because the Stokes {\it Q} and {\it U} components give the polarized intensity, $\sqrt{{\it Q}^2 + {\it U}^2}$. This bias has a particularly significant effect in measurements with a low signal-to-noise ratio.
We therefore debiased the PI map as ${\rm PI}=\sqrt{{\it Q}^2 + {\it U}^2}-\sigma_{\rm PI}$ as described by \citet{vai06,hul15}, where $\sigma_{\rm PI}$ is the rms noise level.
The rms noise levels of the PI emission are $\sigma_{\rm PI}=62.9$ $\mu$Jy beam$^{-1}$ for the 0.87 mm and $\sigma_{\rm PI}=21.8$ $\mu$Jy beam$^{-1}$ for the 3.1 mm, respectively.
The polarization vectors are overlaid with the polarized intensity images where the emission is above 3$\sigma_{\rm PI}$.

For the Stokes {\it I} images, the total flux density is estimated to be $890\pm10$ mJy for 0.87 mm and $48.3\pm0.1$ mJy for 3.1 mm, respectively, by integrating the emission within the disk region on the image.
The peak position is (R.A. Dec) = (4h27m4.700s, $26^\circ6'15\farcs70$) with an intensity of $197.8\pm0.2$ mJy beam$^{-1}$ corresponding to $53.3\pm0.1$ K in the brightness temperature for 0.87 mm.
The peak position is (R.A. Dec) = (4h27m4.700s, $26^\circ6'15\farcs69$) with an intensity of $21.22\pm0.03$ mJy beam$^{-1}$ corresponding to $51.0\pm0.1$ K in the brightness temperature for 3.1 mm.

The polarized intensity images at 0.87 mm and 3.1 mm show different morphologies.
The 0.87 mm polarized intensity is mainly detected in the inner region of the disk, while the 3.1 mm polarized intensity shows a whole with a small detection in the center.
The polarization vectors at 0.87 mm are mainly parallel to the disk minor axis in the inner region ($r\lesssim40$ au) and imply the azimuthal direction in the outer region of the disk ($r\gtrsim40$ au), which is also suggested by \citet{bac18,pod20}.
In contrast, the polarization vectors at 3.1 mm are mainly azimuthal in the outer region of the disk. In the central region, however, the polarization vectors are parallel to the disk minor axis.
These different patterns of the dust polarization are quite consistent with the case of HL Tau as shown by \citet{kat17,ste17}.
In the HL Tau disk, \citet{yan19,mor21} suggest that the morphology at 0.87 mm is consistent with the expectation for self-scattering, while that at 3.1 mm shows grains aligned with the radiation anisotropy \citep{taz17} or gas flow \citep{kat19}. 
The DG Tau disk would be the same case with HL Tau.

\begin{figure}[htbp]
\begin{center}
\includegraphics[width=8.cm,bb=0 0 1014 2250]{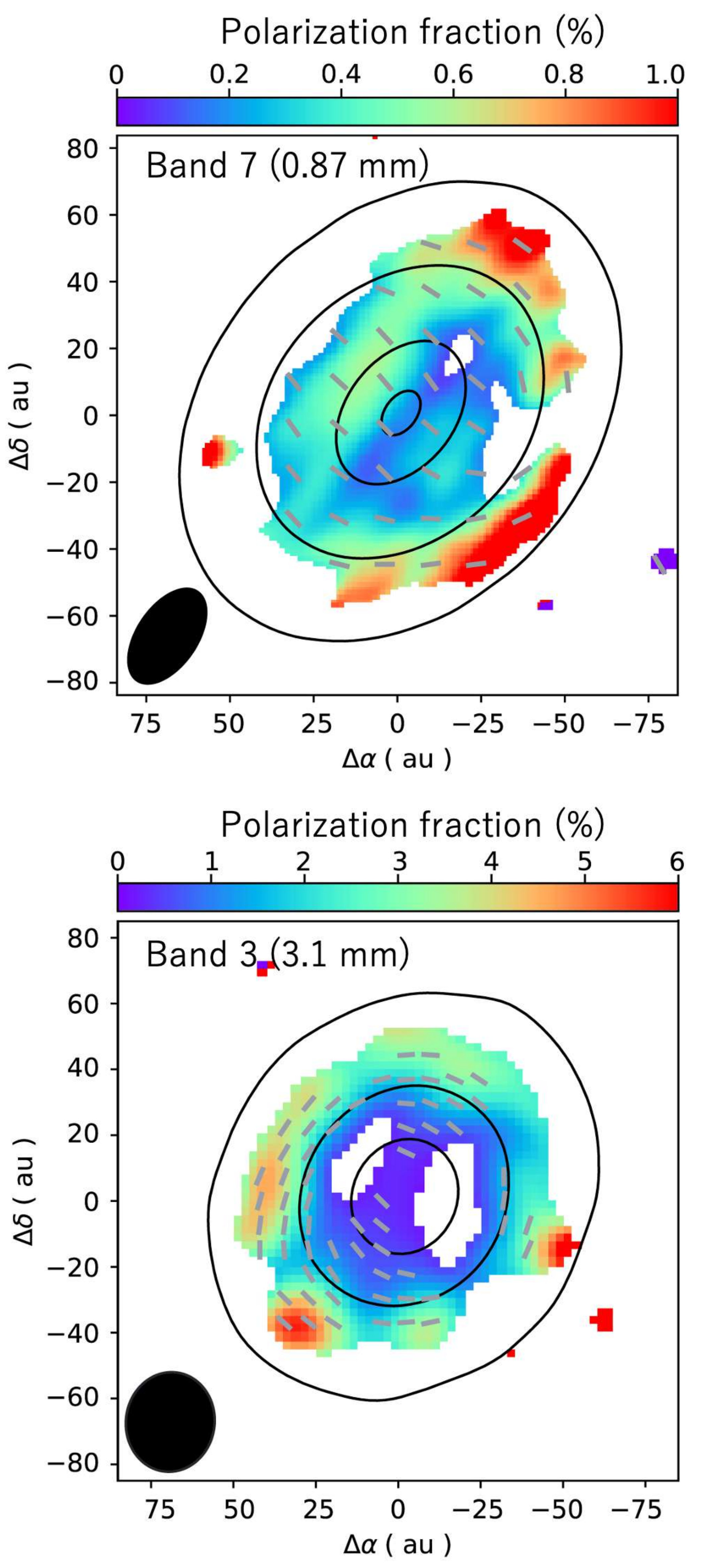}
\end{center}
\caption{ALMA images of polarization fraction (PI/Stokes {\it I}) for wavelengths of 0.87 mm (top panel) and 3.1 mm (bottom panel). The gray bars show the direction of the polarization vectors. The contours indicate the Stokes {\it I} emission same as Figure \ref{polari_image}.
The bottom left corner in each panel shows the beam sizes.
}
\label{polari_frac_image}
\end{figure}

Figure \ref{polari_frac_image} shows the polarization fraction (PI/Stokes {\it I}) for each wavelength.
At 0.87 mm there is an asymmetry along the disk minor axis, with the northeast more polarized than the southwest, and a sudden drop toward the center.
This asymmetry along the disk minor axis is interpreted as a flared disk based on the self-scattering theory \citep{yan16,bac18}.
The polarization fraction reaches $\gtrsim1\%$ in the outer region of the disk, where the polarization vectors become azimuthal, consistent with polarization mechanisms caused by thermal emission of radiatively or mechanically aligned non-spherical grains rather than the self-scattering.
For 3.1 mm, the polarization fraction reaches as high as $3-6\%$ in the outer region of the disk ($r\approx50-60$ au) with the azimuthal polarization vectors, which is the same as for the HL Tau disk \citep{kat17,ste17,mor21}.
The high polarization fraction ($\gtrsim1-2\%$) with the azimuthal polarization vectors is consistent with the grain alignment mechanisms for dust polarization.

In summary, the polarization fraction is derived to be  $\approx0.6\%$ at 0.87 mm and $\approx0.3\%$ at 3.1 mm, respectively, in the central region with the polarization vectors parallel to the disk minor axis, suggesting that the polarization is caused by the self-scattering.
In contrast, the polarization fraction becomes as high as $\approx1-6\%$ in the outer region of the disk ($r\gtrsim50$ au) at both 0.87 mm and 3.1 mm.
In addition, the polarization vectors are azimuthal. These results in the outer region are consistent with the expectations of the grain alignment theories rather than the self-scattering.

\subsection{Radial Intensity Profiles}

The dependence of dust thermal continuum emission on frequency (wavelength) is a reliable method for estimating dust size \citep[e.g.,][]{dra06,bir18}.
Therefore, by performing a multi-band analysis of the spectral energy distribution (SED) at millimeter wavelengths across multiple bands, it is possible to derive information about the dust surface density, temperature, and grain size, as discussed in Section \ref{sec:analysis}.

Before the SED analysis, CLEAN images were acquired using Briggs weighting with a robust parameter of $-2$ for each wavelength. We employed the same image size and cell size with the same uvdistance ($30k-1800k\lambda$).
All of the reconstructed CLEAN images were convolved to have a beam with an FWHM of $0\farcs197\times0\farcs147$, which is the largest beam among the CLEAN images, to directly compare the intensity distributions with minimizing the beam dilution effect.
The noise levels of the CLEAN maps are 320, 52, and 64 $\mu$Jy beam$^{-1}$ for 0.87 mm, 1.3 mm, and 3.1 mm, respectively.
The total flux densities are $877\pm2$ mJy, $360\pm1$ mJy, and $48.1\pm0.2$ mJy, respectively.
These total flux densities are similar to the CLEAN images created with a robust parameter of $-0.5$. 
 In addition, the total flux of the 1.3 mm emission is consistent with the previous CARMA observations that were observed with shorter baseline range of $\gtrsim21$ m corresponding to $\gtrsim16$ k$\lambda$ in uv distance.
 The similarity of the total flux densities regardless of uv weighting parameters implies that most of the flux is recovered by the observations.
  Even if an extended emission is not covered by our observations,  such extended emission would be dominated by the envelope component, which can be ignored in this study because we focus on the disk structure rather than the envelope.

\begin{figure}[htbp]
\begin{center}
\includegraphics[width=8.cm,bb=0 0 1508 2163]{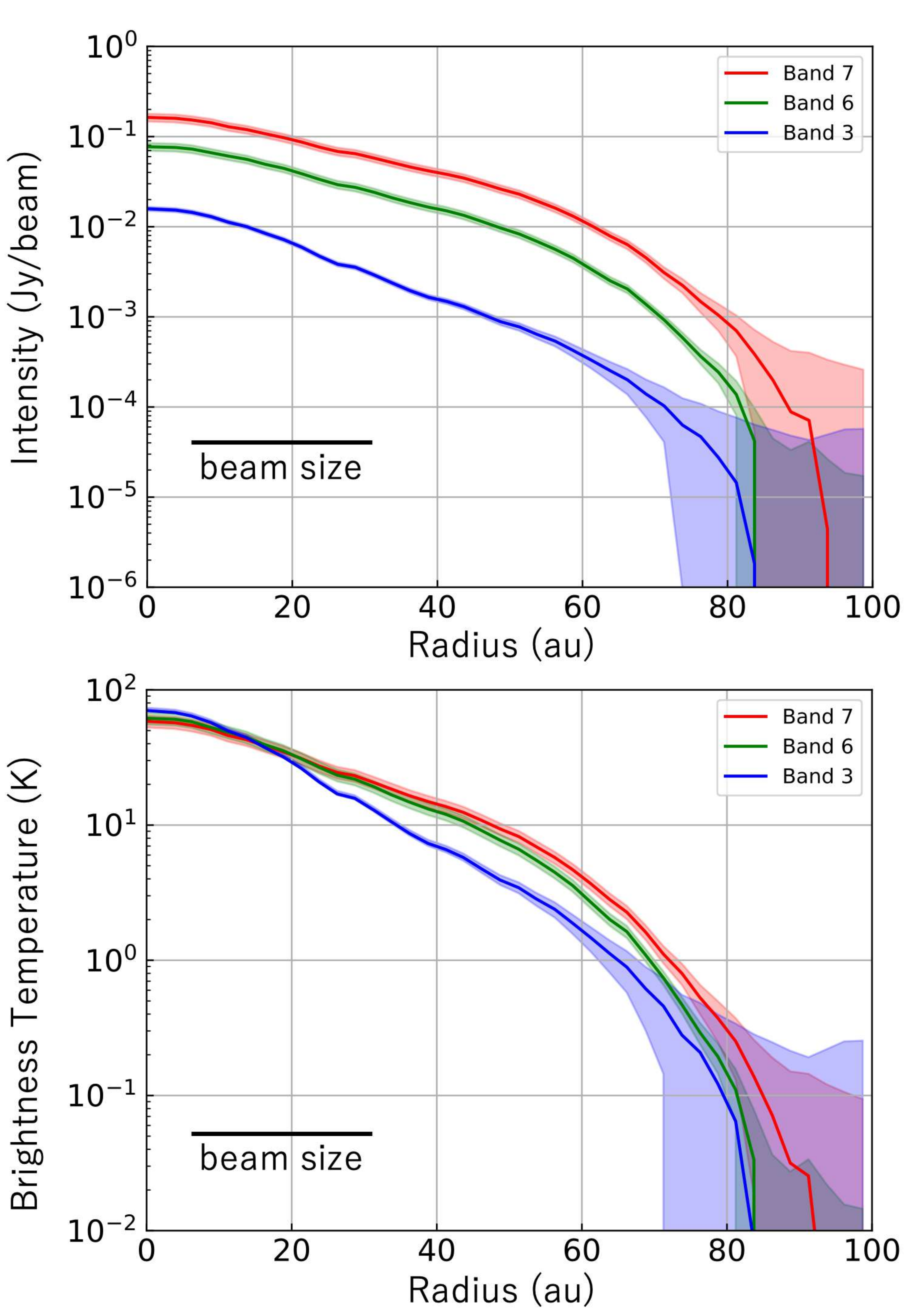}
\end{center}
\caption{Deprojected radial profiles at ALMA Bands 7, 6, and 3. The intensity is shown as Jy beam$^{-1}$ (top panel) and as the brightness temperature assuming the Rayleigh-Jeans approximation (bottom panel).
The shadowed regions correspond to the $1\sigma$ errors including the rms noise levels and flux uncertainties of 10\% for Bands 7 and 6, and 5\% for Band 3.
}
\label{radial_profile}
\end{figure}

Figure \ref{radial_profile} shows the azimuthally averaged intensity profile at 0.87 mm, 1.3 mm, and 3.1 mm.
The intensity of the upper panel is the Jy beam$^{-1}$, while that of the lower panel is set to be the brightness temperature, assuming the Rayleigh-Jeans approximation.
We plot the brightness temperature profile to see the intensity dependence on the wavelength.
The central region indicates almost the same temperature of $60-70$ K. The highest temperature of $70\pm4$ is found at the 3.1 mm emission, while the 1.3 mm and 0.87 mm  emission show a temperature of $62\pm6$ K and $59\pm6$ K, respectively.
Note that the error is dominated by the uncertainty in the flux calibration.
The similar temperature suggests that the emission is optically thick at the observed wavelengths, even in the 3.1 mm emission.
Indeed, a recent survey of SED analysis from optical to radio wavelengths suggests that most disks are optically thick at millimeter wavelengths, even up to 3 mm in the Lupus star-forming region \citep{xin23}.
The temperature of the 3.1 mm emission becomes lower than that of the 0.87 mm and 1.3 mm emission toward the outer disk region, suggesting that the 3.1 mm emission becomes optically thinner in the outer disk region.

\section{Triple ALMA bands analysis}\label{sec:analysis}

In this section, we characterize the DG Tau disk using the spatially resolved ALMA observations at Bands 3, 6, and 7.

\subsection{dust continuum data}\label{sec:sed}

To determine the structure of the dust disk, we perform the multi-band analysis method previously employed in \citet{mac21,sie21,ued22,gui22,zha23}.  This approach attempts to reproduce the intensity at each radius using a set of parameters, including the dust temperature $T$, dust surface density $\Sigma_{\rm d}$, and the maximum dust radius $a_{\rm max}$.
Given these three parameters, the intensity is calculated as \citep{sie19,car19}
\begin{eqnarray}
I_{\rm \nu}=B_{\rm \nu}(T)\left\{ 1-\exp{\left(-\frac{\tau_{\nu}}{\mu}\right)} +\omega_{\nu} F(\tau_{\nu},\omega_{\nu})\right\},
\label{eq:intensity}
\end{eqnarray}
where $B_{\nu}(T)$ is the Planck function with temperature $T$, $\tau_{\nu}$ is the vertical extinction optical depth, $\omega_{\nu}$ is the effective scattering albedo and $\mu\equiv\cos{i}$ represents the effect of the disk inclination.
The scattering effect results from $F(\tau_{\nu},\omega_{\nu})$, which is given by
\begin{eqnarray}
F(\tau_{\nu},\omega_{\nu})=\frac{f_{1}(\tau_{\nu},\omega_{\nu})+f_{2}(\tau_{\nu},\omega_{\nu})}{\exp{(-\sqrt{3}\epsilon_{\nu}\tau_{\nu})}(\epsilon_{\nu}-1)-(\epsilon_{\nu}+1)}
\label{eq:Ffactor}
\end{eqnarray}
where
\begin{eqnarray}
f_{1}(\tau_{\nu},\omega_{\nu})=\frac{ 1-\exp{\left\{ -(\sqrt{3}\epsilon_{\nu}+1/\mu)\tau_{\nu}\right\}} }{\sqrt{3}\epsilon_{\nu}\mu+1}
\end{eqnarray}
and
\begin{eqnarray}
f_{2}(\tau_{\nu},\omega_{\nu})=\frac{\exp{\left(-\tau_{\nu}/\mu\right)}-\exp{(-\sqrt{3}\epsilon_{\nu}\tau_{\nu})}}{\sqrt{3}\epsilon_{\nu}\mu-1},
\end{eqnarray}
with $\epsilon_{\nu}\equiv\sqrt{1-\omega_{\nu}}$.
We vary $T$, $\Sigma_{\rm d}$, and $a_{\rm max}$ in a 1000 $\times$ 600 $\times$ 115 grid, respectively.
$T$ is uniformly spaced from 1 to 200 K. $\Sigma_{\rm d}$, and $a_{\rm max}$ are varied in logarithmic space from $10^{-3}$ to $10^2$ g cm$^{-2}$ and from 10$^{-3}$ to 10 cm, respectively.
For the dust opacities, we use the DSHARP model \citep{bir18} that considers a compact spherical dust
with a size distribution ranging from 0.01 $\mu$m to $a_{\rm max}$ with a power-law size distribution of $q=-3.5$.

Once we obtain the calculated intensities at each wavelength from Equation (\ref{eq:intensity}), we evaluate their posterior probability distributions using the chi-square statistic as
\begin{eqnarray}
\chi^{2}=\sum_{i}\left(\frac{I_{{\rm obs},i}-I_{{\rm m},i}}{\sigma_{i}}\right)^{2},
\label{eq:chi}
\end{eqnarray}
where $I_{{\rm obs},i}$ and $I_{{\rm m},i}$ are the observed and calculated intensity at frequency $i$, respectively.
The uncertainty in the observed intensity at frequency $i$, $\sigma_{i}$, is given by
\begin{eqnarray}
\sigma_{\rm i}^{2} = \Delta I_{{\rm obs},i}^{2} + (\delta_{\rm i} I_{{\rm obs},i})^{2},
\end{eqnarray}
where $\Delta I_{{\rm obs}, i}$ is the standard deviation of the azimuthally averaged intensity at frequency $i$ and $\delta_{i}$ represents the absolute flux calibration error at frequency $i$.
We set $\delta_{i}$ to 5\% for ALMA Band 4 and 10\% for Bands 6 and 7, following the ALMA observing guide.

\begin{figure*}[htbp]
\begin{center}
\includegraphics[width=17.cm,bb=0 0 1900 2250]{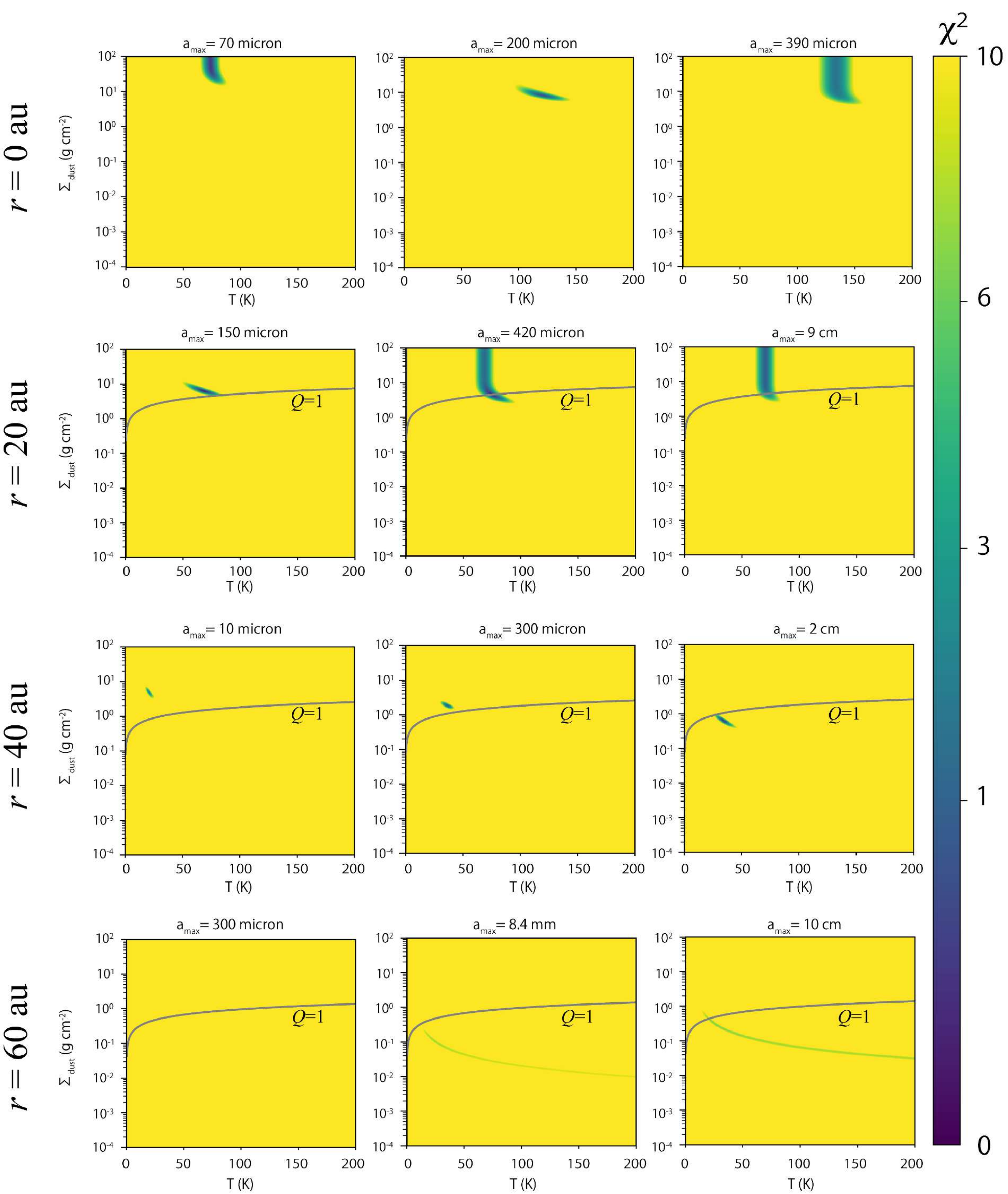}
\end{center}
\caption{The $\chi^2$ distributions of the SED fitting (Equation \ref{eq:chi}) in the parameter spaces of the dust surface density ($\Sigma_{\rm d}$) and the dust temperature ($T$) at a given dust size ($a_{\rm max}$). 
The each row shows the $\chi^2$ distributions where the disk radius of $r=0$, $20$, $40$, and $60$ au.
}
\label{xi_fit}
\end{figure*}

\begin{figure}[htbp]
\begin{center}
\includegraphics[width=8.cm,bb=0 0 2445 1765]{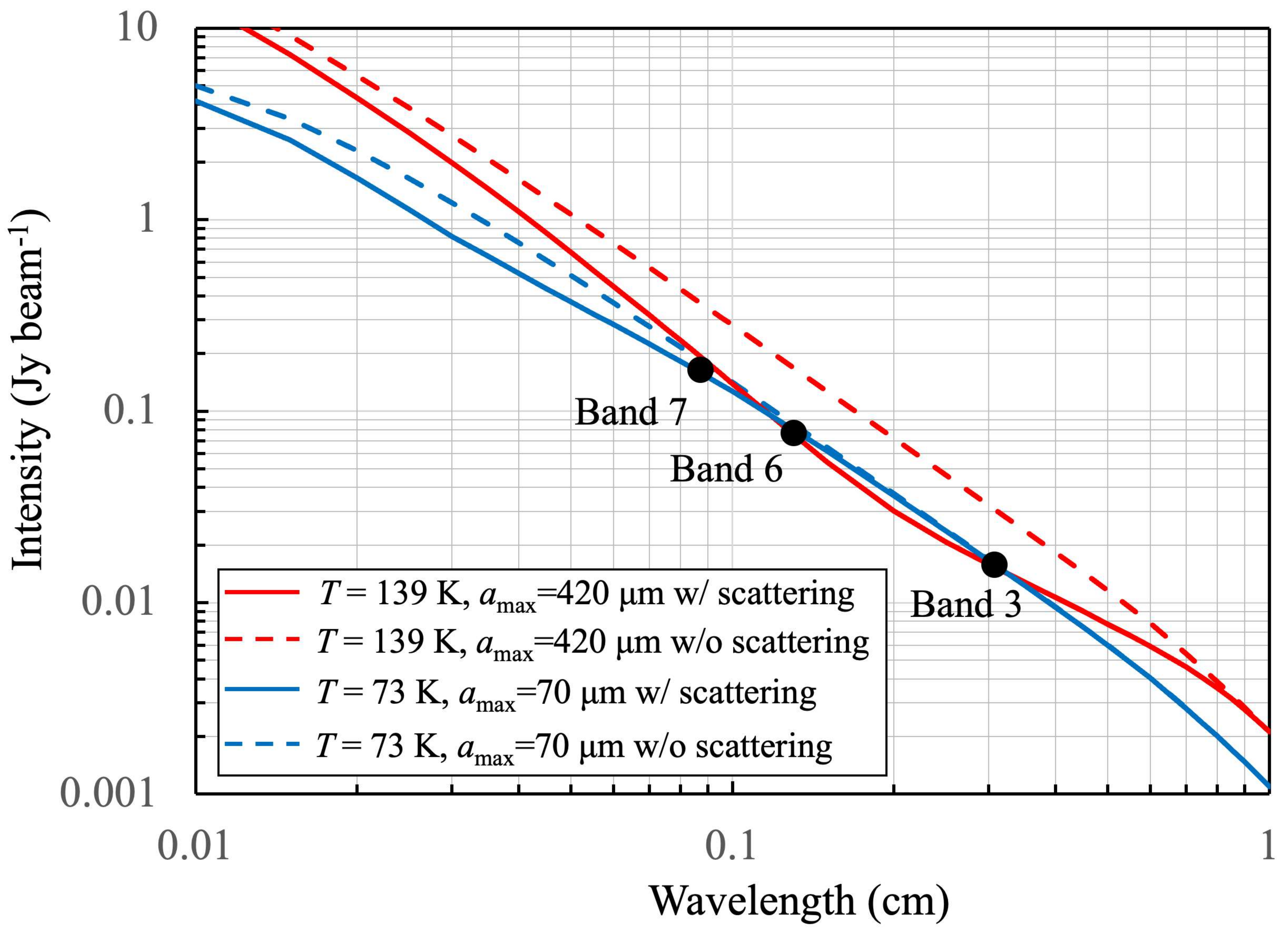}
\end{center}
\caption{SED examples of two of the best SED fitting results at the disk center of $r=0$ au.
The case of $T=139$ K and $a_{\rm max}=420$ $\mu$m is shown as the red line, while the case of $T=73$ K and $a_{\rm max}=70$ $\mu$m is shown as the blue line. Both cases assume $\Sigma_{\rm d}=100$ g cm$^{-2}$ as a representative of optically thick conditions.
The dashed lines indicate the cases without the scattering effect.
}
\label{sed}
\end{figure}

Figure \ref{xi_fit} shows some examples of the chi-square statistics at each radius with 5 au bins.
In each panel, the parameters of $\Sigma_{\rm d}$ and $T$ are shown in the vertical and horizontal axes, respectively, with a given $a_{\rm max}$.
As shown in the figure, there are several parameter spaces in different $a_{\rm max}$ for fitting to the observed intensities. For example, at the disk center of $r=0$ au, a surface density larger than $\Sigma_d\gtrsim20$ g cm$^{-2}$ can reproduce the observed intensities with a $a_{\rm max}$ range of $a_{\rm max}\sim10-400$ $\mu$m and a $T$ range of $T\sim60-150$ K.
To confirm that these different parameters actually reproduce the observed intensities, we show the SED for the $a_{\rm max}=70$ $\mu$m and $a_{\rm max}=400$ $\mu$m grain solutions at the disk center ($r=0$ au) in Figure \ref{sed}.
The blue and red lines show the models of the $a_{\rm max}=70$ $\mu$m grains at $T=73$ K and the  $a_{\rm max}=400$ $\mu$m grains with $T=139$ K, respectively. Both models assume $\Sigma_d=100$ g cm$^{-2}$ (a representation of an optically thick case).
We also show the cases without the scattering effect to see the contribution of the scattering to the intensities as the blue and red dashed lines.
We confirm that the emission becomes fainter if the dust scattering is considered as shown by \citet{soo17,liu19,zhu19}. 
 The calculations with taking into account the scattering effect clearly reproduce the observed intensities.
Therefore, our multi-band analysis works well, and the emission is quite optically thick ($\tau>>1$) even at a wavelength of 3.1 mm.

In Figure \ref{xi_fit}, we also plot the Toomre {\it Q} value as the threshold at which the disk is gravitationally unstable, derived from the criterion \citep{too64}:
\begin{eqnarray}
Q\equiv\frac{c_{\rm s}\Omega_{\rm K}}{\pi G \Sigma_{\rm g}} =1,
\label{eq:Q}
\end{eqnarray}
where $c_{\rm s}$ is the sound speed of the gas, $\Omega_{\rm K}$ is the Keplerian rotational frequency, $G$ is the gravitational constant and $\Sigma_{\rm g}$ is the gas surface density.
For simplicity, we assume a constant dust-to-gas mass ratio of 0.01.
The stellar mass of DG Tau is set to $0.7M_{\odot}$.
Interestingly, we found that almost all of the possible solutions of the dust models are  above the critical value where $Q=1$ at $r=20$ au.
This suggests that the DG Tau disk needs to be gravitationally unstable unless the dust-to-gas mass ratio becomes higher.
Based on these results, we discuss the dust enrichment in Section \ref{sec:d2g}.

\begin{figure*}[htbp]
\begin{center}
\includegraphics[width=17.cm,bb=0 0 4464 1603]{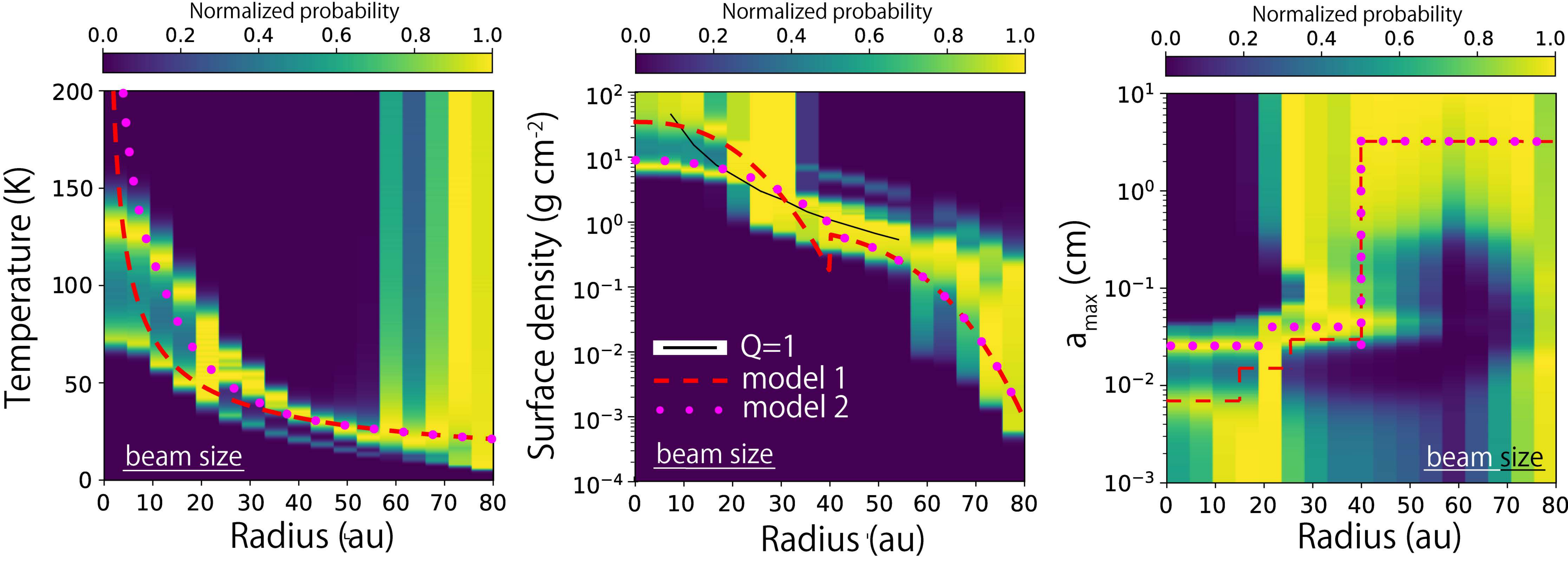}
\end{center}
\caption{Posterior probability distributions of the fitting parameters for the multiband observations. The marginalized posterior probability is normalized by the maximum at each radius.
The black  line in the middle panel shows the dust surface density profile where the disk is gravitationally unstable (i.e., $Q = 1$), assuming a gas-to-dust ratio of 100 and the higher temperature in the highest probabilities.
The red dashed lines and pink dotted lines show the models discussed in Section \ref{sec:model1} and \ref{sec:model2}, respectively.
}
\label{probability}
\end{figure*}

The posterior probability distributions of the fitting parameters for the multiband observations at each disk radius can be computed as
\begin{eqnarray}
P(I_{\rm B3},I_{\rm B6},I_{\rm B7}|T,\Sigma_{\rm d},a_{\rm max}) \propto \exp{\left(-\frac{\chi^{2}}{2}\right)}.
\end{eqnarray}
Figure \ref{probability} shows the posterior probability distributions of each of the fitting parameters.
There are certain ranges of the possible solutions for the fitting parameters as shown in Figure \ref{xi_fit}.
Within a 20 au radius, the two best solutions ($p\gtrsim0.9$) can be seen: higher temperature ($T\approx 140$ K) with relatively larger dust grains  ($a_{\rm max}\approx 240$ $\mu$m) and lower temperature  ($T\approx 70$ K) with smaller dust grains  ($a_{\rm max}\approx 70$ $\mu$m).
These solutions require a lower surface densities ($\Sigma_{\rm d}\approx10$ g cm$^{-2}$) and  higher surface densities ($\Sigma_{\rm d}\approx40$ g cm$^{-2}$), respectively.
Although the solutions have these certain ranges for the high probability ($p\gtrsim0.9$), we can constrain the physical conditions from the probability distributions.
We found that the surface density, $\Sigma_{\rm d}$, needs to be larger than $\gtrsim7$ g cm$^{-2}$ in the central region, suggesting that the 3.1 mm emission is optically thick because the extinction optical depth is $\tau_{\nu}\gtrsim1.8$ at 3.1 mm.
Within a radius of 20 au, $a_{\rm max}$ needs to be less than $400$ $\mu$m, while it becomes as large as 1 cm in the outer radius.
These different grain size distributions are discussed in Section \ref{sec:origin}.

The critical surface density where $Q=1$ is also shown in the probability distribution of $\Sigma_{\rm d}$ as the black line.
Here, we use the higher temperature value of the probability solution to derive the ${\it} Q$ value as an upper limit.
We found that the most possible solution of the surface density is above the critical surface density of $Q=1$ at a 20 au radius.
This suggests that DG Tau needs to be gravitationally unstable unless the dust-to-gas mass ratio becomes larger. We discuss the dust-to-gas mass ratio in Section \ref{sec:d2g}.

\subsection{Comparison between models and polarization data}

In the previous subsection, we performed the multi-band analysis by using the 0.87 mm, 1.3 mm, and 3.1 mm dust continuum emission and showed the possible solutions of the temperature, surface density, and grain size distributions.
In this subsection, we investigate whether the solutions can agree with the polarization observations by performing radiative transfer calculations, RADMC3D \citep{dul12}.
Here, we consider two models as mentioned above: lower temperature  ($T\approx 70$ K) with smaller dust grains  ($a_{\rm max}\approx 70$ $\mu$m) and higher temperature ($T\approx 140$ K) with relatively larger dust grains  ($a_{\rm max}\approx 240$ $\mu$m) in the central region.
In our calculations, spherical dust grains are considered as in the previous subsection, which means that our calculations do not reproduce the polarization due to the alignment of non-spherical dust grains. The motivation here is not to explain all of the polarization patterns seen in the observations, but to find to what extent the scattered polarization alone can explain the observations.

\subsubsection{Smaller dust with a single power-law temperature model}\label{sec:model1}

In this subsection, we calculate the smaller dust grain model ($a_{\rm max}\approx 70$ $\mu$m in the central region).
According to Figure \ref{probability}, the grain size increases with radius.
Based on Figure \ref{probability}, we assume the grain size distributions as follows
\begin{equation}
a_{\rm max} = \begin{cases}
70\ {\rm \mu m}, & r \leq 15 \ {\rm au}, \\
160\ {\rm \mu m}, & 15 < r \leq 25 \ {\rm au}, \\
300\ {\rm \mu m}, & 25 < r \leq 40 \ {\rm au}, \\
1\ {\rm mm},\ 3\ {\rm mm},\  1\ {\rm cm} \ {\rm or}\  3\ {\rm cm} & r > 40 \ {\rm au}. \\
\end{cases}
\label{amax}
\end{equation}
For the different maximum grain sizes in the outer region ($ r > 40 \ {\rm au}$), we create each model by varying the surface density.

For the dust grains with $a_{\rm max}=70$ $\mu$m, the brightness temperature reflects the dust temperature because the emission is optically thick and the scattering effect for the intensity is negligible.
The radial profile of the 1.3 mm dust continuum emission shows a power law of $-0.6$ within a radius of $\approx12$ au in the brightness temperature.
Therefore, we assume that the dust temperature is
\begin{equation}
T=294 \Big(\frac{r}{1\ \rm au}\Big)^{-0.6}\ \rm K.
\end{equation}
The assumed temperature of 294 K at a radius of 1 au is determined by the brightness temperature of the 1.3 mm dust continuum emission.

The dust surface density is assumed to have two exponential power laws to reproduce the change of the intensity slope around $r\approx45$ au found in the 1.3 mm emission. The peak surface density is inferred to be $\gtrsim35$ g cm$^{-2}$ derived from the multi-band analysis. The surface density profile is then given by
\begin{equation}
\Sigma_{\rm d} = \begin{cases}
35 \exp\left[-\Big( \frac{r}{r_{\rm c1}} \Big)^{-\gamma_1}\right]\  {\rm g~cm^{-2}}, & r \leq 40 \ {\rm au}, \\\\
\Sigma_2 \exp\left[-\Big( \frac{r}{r_{\rm c2}} \Big)^{-\gamma_2}\right]\  {\rm g~cm^{-2}}, & r > 40 \ {\rm au}.
\end{cases}
\label{eq:ts}
\end{equation}
With these assumptions, we try to fit the observed intensity profiles of the 0.87 mm, 1.3 mm, and 3.1 mm emission by searching for the parameters of $r_{\rm c1}$, $\gamma_{\rm 1}$, $\Sigma_2$, $r_{\rm c2}$, and $\gamma_2$. 
 The best fit values of these parameters are listed in Table \ref{tbl:model1}.

\begin{table}[!ht]
\centering
\vspace{0mm}
\caption{best fit parameters of smaller dust with a single power-law temperature model}
\begin{tabular}{lcl}
\hline
\hline
fitting parameter & value \\
\hline
$r_{\rm c1}$ &  20 au &\\
$\gamma_{\rm 1}$  & 2.5 & \\
$\Sigma_2$ & 0.38 &for $a_{\rm max}=1$ mm \\
                    &0.42  &for $a_{\rm max}=3$ mm\\
                    &  0.59 &for $a_{\rm max}=1$ cm \\
                    &  0.90 &for  $a_{\rm max}=3$ cm \\
$r_{\rm c2}$ & 52 au &\\
$\gamma_2$ & 4.4 &\\
\hline
\end{tabular}\\
\label{tbl:model1}
\end{table}

For calculating the intensity, the dust scale height ($H_{\rm d}$) also needs to be assumed.
In this calculation, we assume that $H_{\rm d}$ is one-third of the gas scale height ($H_{\rm g}$).
The gas scale height is determined by
\begin{equation}
H_{\rm g}=c_s/\Omega_{\rm K},
\label{eq:scale_height}
\end{equation}
where $c_s$ is the sound speed and $\Omega_{\rm K} = \sqrt{GM_\star/r^3} = 2.0\times 10^{-7} (r/1~{\rm au})^{-3/2}(M_\star/M_\odot)^{1/2}~{\rm s^{-1}}$
is the Keplerian frequency with $G$, $M_\star$ being the gravitational constant and central stellar mass, respectively.
Note that we investigate the dust scale height and discuss the dust settling in Section \ref{sec:scale_height}.

\begin{figure*}[htbp]
\begin{center}
\includegraphics[width=17.cm,bb=0 0 3920 976]{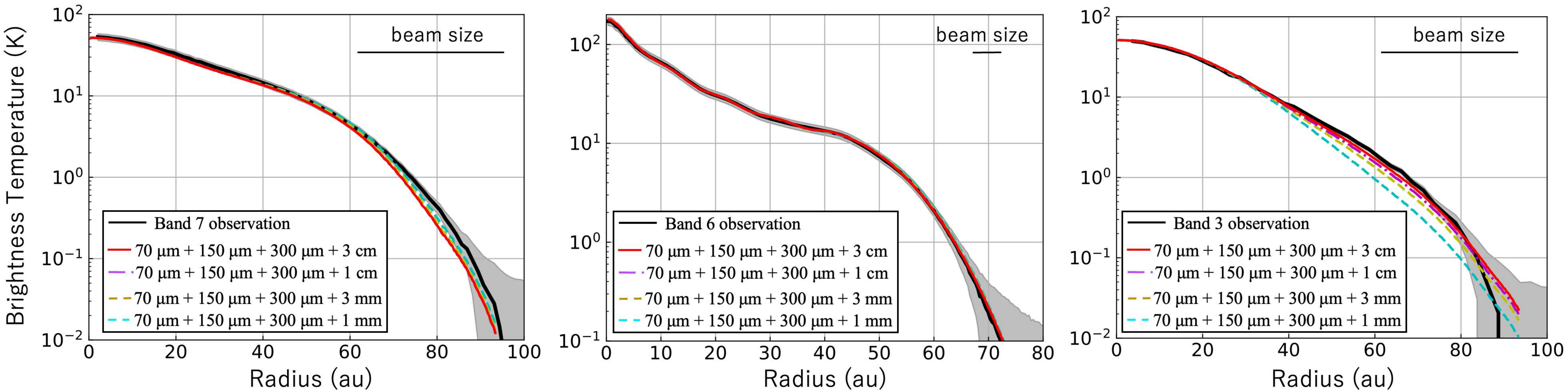}
\end{center}
\caption{Comparison of radial profiles between the observed intensities and model intensities.  The model is described in Section \ref{sec:model1}. The black solid lines indicate the observed intensities with uncertainties of the rms noise levels and the flux calibration errors, shown as the shaded region.
The red solid, cyan dashed, and purple dashed lines show the model intensities with dust sizes of 1 cm, 3 mm, and 1 mm in the outer region of the disk ($r\geq40$ au), respectively.
}
\label{model1_profile}
\end{figure*}

Figure \ref{model1_profile} shows the comparisons of the radial profiles of the intensities between the observations and the radiative transfer calculations of our best-fit model. 
We can clearly see that the model reproduces the observed intensity well within the 3$\sigma$ errors.
The slight wiggle profile found in the 1.3 mm emission is naturally reproduced as the grain sizes in the radius change due to the scattering effect.
However, the model with $a_{\rm max}=1$ mm in the outer region of the disk indicates a slightly lower intensity than the observed 3.1 mm emission, while the models with  $a_{\rm max}=3$ mm,  $a_{\rm max}=1$ cm, and  $a_{\rm max}=3$ cm reproduce the observed intensity.
These models suggest that the maximum grain size is likely to exceed 3 mm in the outer region of the disk although the case of $a_{\rm max}=3$ cm is the best fit to the observations, which is consistent with the results of the multi-band analysis (shown in Figure \ref{probability}).
The best-fit model is also shown in Figure \ref{probability} as red dashed lines (model 1).
Note that this fitting approach does not apply to the statistical survey. 
It may be possible to use other parameters to reproduce the observations.
However, the results follow the most possible solutions of the probability distributions, suggesting that the model is reliable.

\begin{figure*}[htbp]
\begin{center}
\includegraphics[width=17.cm,bb=0 0 4729 5973]{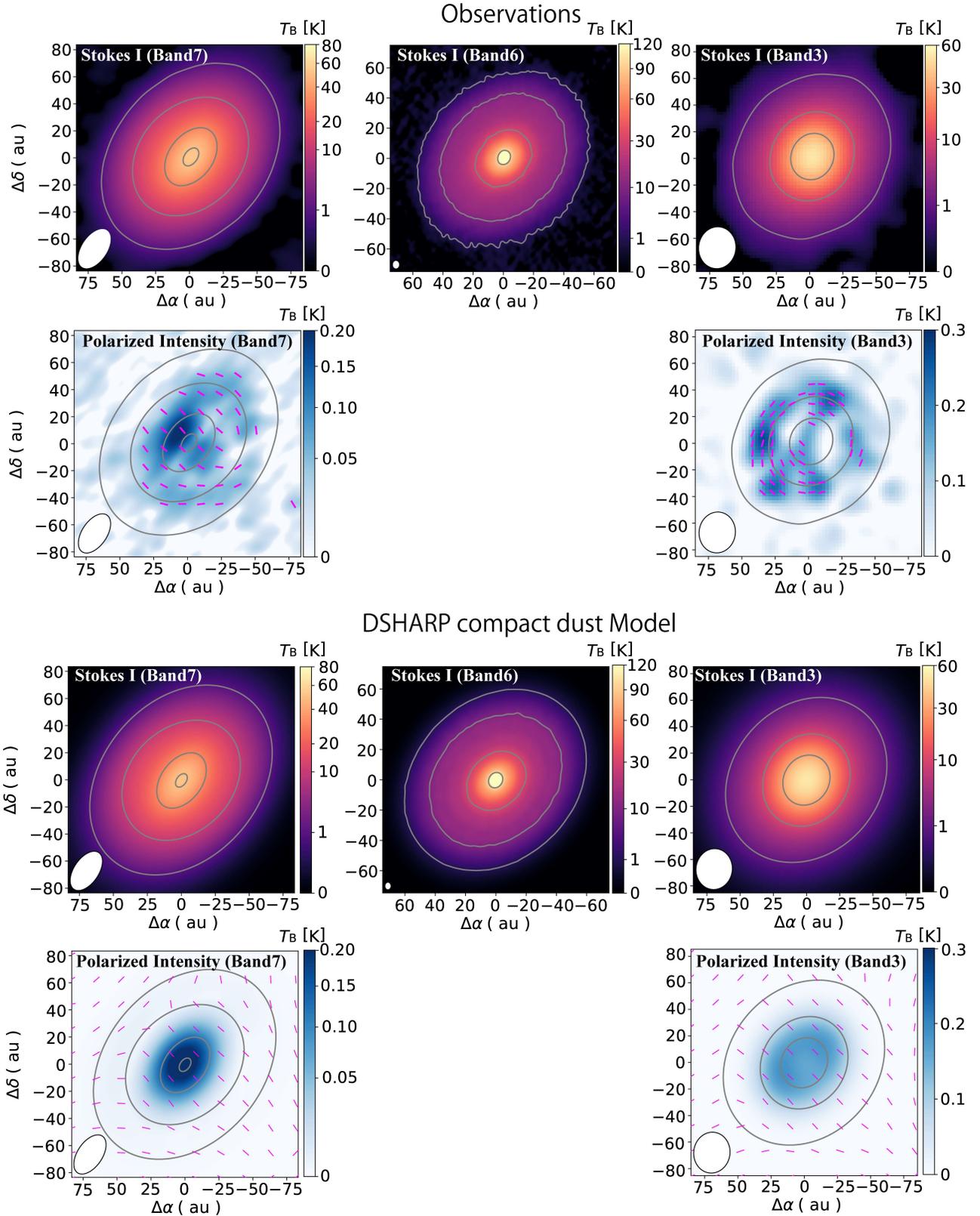}
\end{center}
\caption{Model images at wavelengths of 0.87 mm, 1.3 mm, and 3.1 mm derived by RADMC3D \citep{dul12} from the model described in Section \ref{sec:model1}.  The PI images are also shown at wavelengths of 0.87 mm and 3.1 mm. The contours are the Stokes {\it I} emission of [1, 10, 30, 50] K for 0.87 mm, [1, 10, 30, 100] K for 1.3 mm, and [1, 10, 30] K for 3.1 mm.
}
\label{model1_image}
\end{figure*}

Figure \ref{model1_image} shows continuum and polarization images of the radiative transfer calculations of our best-fit model. 
The continuum images agree well with the observations as described in the radial profile of Figure \ref{model1_profile}.
We found that the polarization images are also similar to the observations.

For the 0.87 mm polarization model, the polarization vectors are parallel to the disk minor axis, and the polarized intensity reaches $T_{\rm B}\approx0.2$ K in the central region, which is consistent with the observations.
However, there are some inconsistencies between the observations and the model. For example, the PI emission is observed in the outer region of the disk ($r\gtrsim40$ au) with the azimuthal polarization vectors, while the model indicates that the PI emission is almost undetectable (${\rm PI}\approx0$ K in $T_{\rm B}$).
This would suggest that the polarization in the outer region may be caused by polarized thermal emission of radiatively or mechanically aligned non-spherical grains rather than by self-scattering since the dust grains with $a_{\rm max}=1$ cm cannot produce the polarization due to the scattering \citep{kat15}.
Around the central region ($r\lesssim20$ au), the observations show the asymmetry of the polarized intensity along the minor axis with the peak of the northeast side, while the model shows the symmetry distribution.
\citet{bac18} suggested that the asymmetry of the polarization may be caused by the flaring of the dust scale height because the radiation anisotropy is enhanced with increasing the scale height \cite[e.g.,][]{yan16,oha19}.

\begin{figure*}[htbp]
\begin{center}
\includegraphics[width=17.cm,bb=0 0 3056 2080]{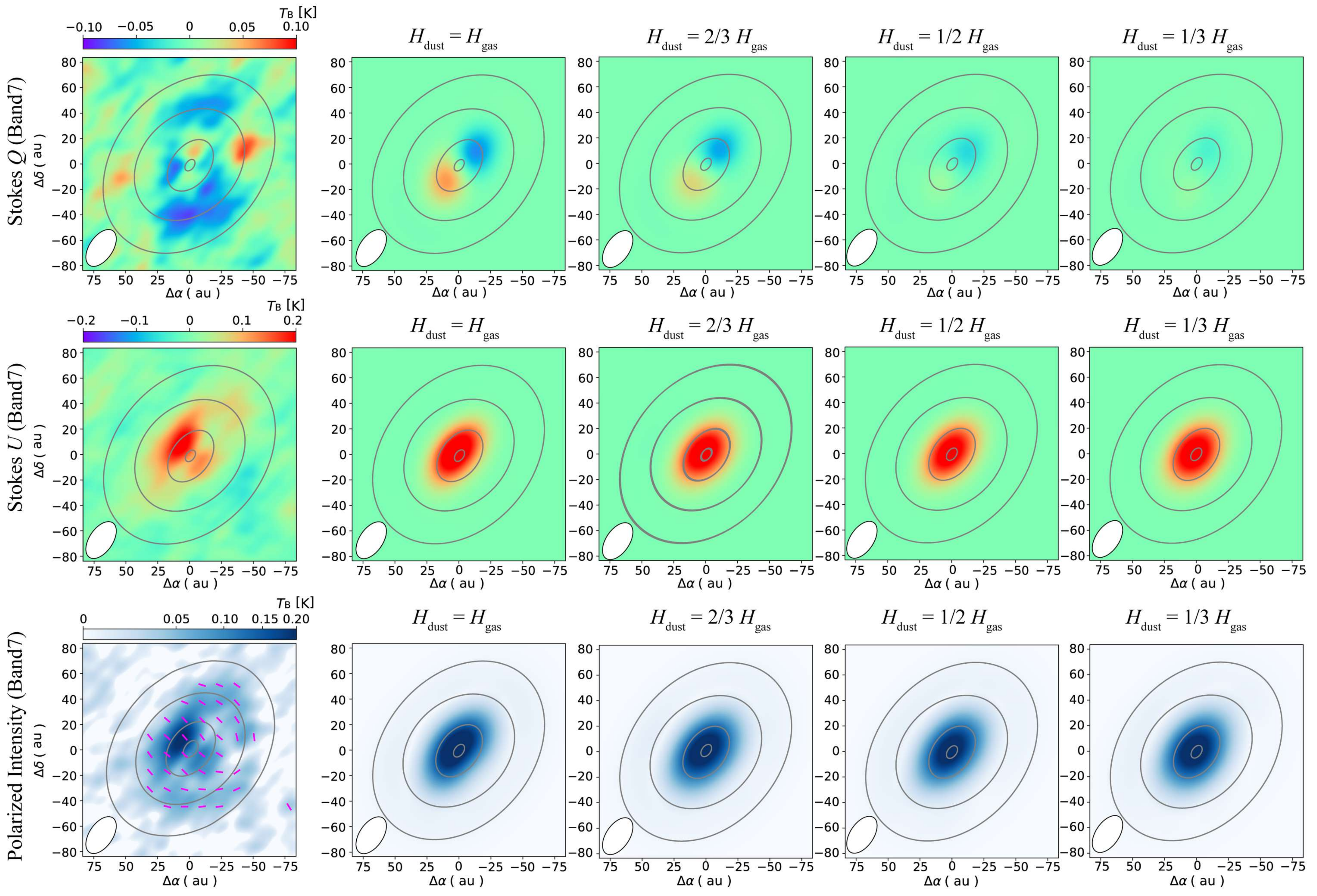}
\end{center}
\caption{Stokes {\it Q} (top row), Stokes {\it U} (middle row), and PI (bottom row) of the model in Section \ref{sec:model1} at a wavelength 0f 0.87 mm is shown for various dust scale heights. The left column displays the observations. The contours are the Stokes {\it I} emission same as Figure \ref{model1_image}.
}
\label{fset}
\end{figure*}

We study the dependence of the polarization on the disk flaring effect by varying the dust scale height in our model.
Figure \ref{fset} shows the comparisons of the observations and models in the Stokes {\it Q}, {\it U}, and PI.
The Stokes {\it Q} emission is found to be enhanced by increasing the dust scale height in the inclined disk as suggested by \citet{oha19}.
Furthermore, the asymmetry in the Stokes {\it U} and PI emission is enhanced: the northeast side is brighter than the southwest.
Although the asymmetry can be confirmed by our model, the steep drop of the Stokes {\it U} and PI emission toward the center and on the disk major axis is not well reproduced by our model.
In the self-scattering theory, the far side of the disk minor axis is expected to be the lowest intensity in PI \citep{yan17}, while the observations show that the far side of the disk minor axis shows the local peak and the lowest intensity in PI is found along the major axis.
This discrepancy may indicate that the polarization is not only produced by self-scattering but also by an additional grain alignment process.
To cancel out the polarization on the disk major axis, the additional polarization requires the radial polarization vectors.
The Stokes {\it Q} image at the 0.87 mm wavelength may give a hint of such radial polarization vectors as the butterfly-shaped structure in the central region. 
In this case, the polarization may be caused by the grain alignment with a toroidal magnetic field similar to the case of the HD 142527 disk \citep{oha18,ste20}.
Our dust size distribution satisfies the condition that the dust size needs to be $\lesssim100$ $\mu$m for the magnetic alignment \citep{laz19,hoa22}.

For the 3.1 mm polarization model, the polarized intensity is generated with a brightness temperature of $T_{\rm B}\approx0.2$ K in the inner region of the disk ($r\lesssim40$ au) because of the suitable grain sizes ($a_{\rm max}\sim0.2-1$ mm) for the self-scattering at 3.1 mm.
The observations also imply a similar brightness temperature in the central position with the polarization vectors parallel to the disk minor axis.
This suggests that our polarization model is consistent with the observations in the inner region.
In contrast, our model predicts a brightness temperature lower than 0.01 K for the polarized intensity in the outer region, because the grain size of $a_{\rm max}\gtrsim3$ mm is too large to produce the polarization for the self-scattering. However, the observations show the brightness temperature as high as $T_{\rm B}\approx0.3$ K with the azimuthal polarization vectors, indicating that the polarization is caused by other mechanisms rather than the self-scattering.

These comparisons of the model and observations suggest that our model matches the observations for the total intensity (Stokes {\it I}) within the $1\sigma$ error.  In addition, the polarization patterns of the model and observations are consistent in the inner region ($r\lesssim40$ au) in terms of the polarized intensity and polarization vectors.
Although there are some inconsistencies between the model and observations for the polarization such as the azimuthal polarization vectors in the outer region of the disk, the other mechanisms such as the dust alignments by radiation or gas flows will contribute to these.
Therefore, we suggest that the proposed grain size distribution is consistent with the polarization induced by the self-scattering.

\subsubsection{Larger dust with a multiple power law temperature model}\label{sec:model2}
Although the previous model already reproduced the observations, we investigate the other possible case of higher temperatures and larger dust grains.
For this case, we assume the following grain size distribution according to Figure \ref{probability}.
\begin{equation}
a_{\rm max} = \begin{cases}
260\ {\rm \mu m}, & r \leq 20 \ {\rm au}, \\
400\ {\rm \mu m}, & 20 < r \leq 40 \ {\rm au}, \\
1\ {\rm cm} & r > 40 \ {\rm au}. \\
\end{cases}
\label{amax}
\end{equation}
The dust size in the outer region ($r>40$ au) is set to be 1 cm, the same as the previous model because the multi-band analysis shows that the dust grains are likely to  be larger than 3 mm.
Therefore, we use the same dust surface density and temperature distributions of the previous model in the outer region  ($r>40$ au) as in the previous model.

The brightness temperature does not reflect the dust temperature in an optically thick condition when the grain size is $a_{\rm max}\approx260$ $\mu$m due to the scattering.
The temperature shows a steeper power law profile than $T\propto r^{-0.6}$ for the larger dust model around $r\approx10-40$ au.
Therefore, we change the temperature distribution as follows.
\begin{equation}
T = \begin{cases}
T_1 \Big(\frac{r}{12\ \rm au}\Big)^{-q_1}\ {\rm K}, & r \leq 20 \ {\rm au}, \\
T_1 \Big(\frac{r}{12\ \rm au}\Big)^{-q_2}\ {\rm K}, & 20 < r \leq 40 \ {\rm au}, \\
T_0 \Big(\frac{r}{1\ \rm au}\Big)^{-0.6}\ {\rm K}, & 40 \ {\rm au} < r. \\
\end{cases}
\label{amax}
\end{equation}
\vspace{0mm}
The temperature distribution is assumed to change smoothly.

The dust surface density is assumed to have a two-exponential power-law, as similar to the previous model:
\begin{equation}
\Sigma_{\rm d} = \begin{cases}
\Sigma_3 \exp\left[-\Big( \frac{r}{r_{\rm c3}} \Big)^{-\gamma_3}\right]\  {\rm g~cm^{-2}}, & r \leq 40 \ {\rm au}, \\\\
\Sigma_2 \exp\left[-\Big( \frac{r}{r_{\rm c2}} \Big)^{-\gamma_2}\right]\  {\rm g~cm^{-2}}, & r > 40 \ {\rm au}.
\end{cases}
\label{eq:ts}
\end{equation}

With these assumptions, we search for the parameters of $T_1$, $q_1$, $q_2$, $\Sigma_3$, and $\gamma_3$ to reproduce the observed intensity profiles of the 0.87 mm, 1.3 mm, and 3.1 mm emission.
 The best fit values of these parameters are listed in Table \ref{tbl:model2}.

\begin{table}[!ht]
\centering
\vspace{0mm}
\caption{best fit parameters of larger dust with a multiple power law temperature model}
\begin{tabular}{lcl}
\hline
\hline
fitting parameter & value \\
\hline
$T_{\rm 1}$ & 102 & \\
$q_{\rm 1}$  & 0.6 & \\
$q_{\rm 2}$  & 0.96  &\\
$\Sigma_3$ & 9 &  \\
$r_{\rm c3}$ & 29 au\\
$\gamma_3$ & 2.5\\
\hline
\end{tabular}\\
\label{tbl:model2}
\end{table}

\begin{figure*}[htbp]
\begin{center}
\includegraphics[width=17.cm,bb=0 0 3926 980]{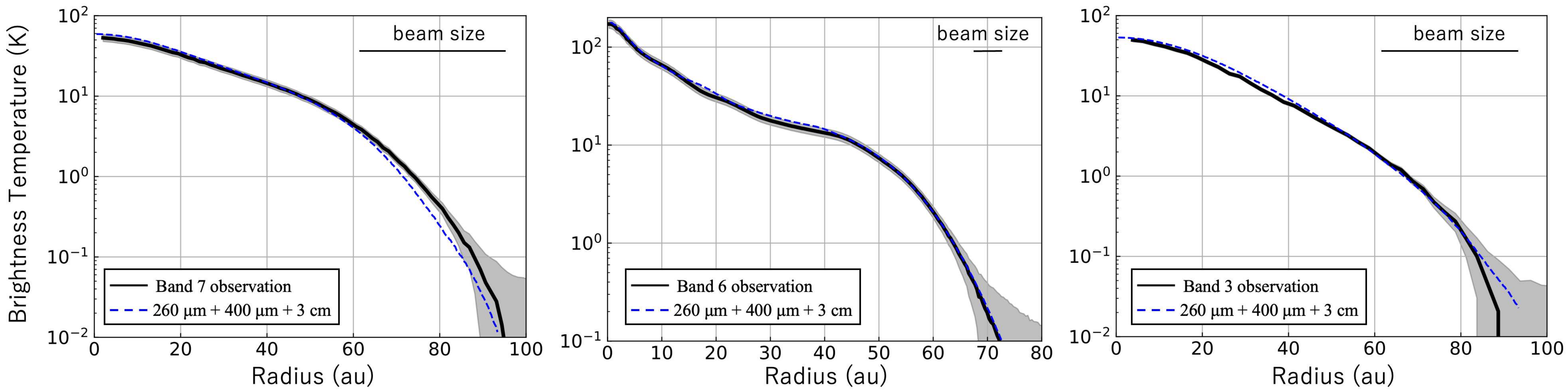}
\end{center}
\caption{Same as Figure \ref{model1_profile} but for the model described in Section \ref{sec:model2}.
}
\label{model2_profile}
\end{figure*}

Figure \ref{model2_profile} shows the comparisons of the radial profiles of the intensities between the observations and the radiative transfer calculations of our best-fit model. Similar to Figure \ref{model1_profile}, this model also reproduces the observed intensity well.

\begin{figure*}[htbp]
\begin{center}
\includegraphics[width=17.cm,bb=0 0 4685 2951]{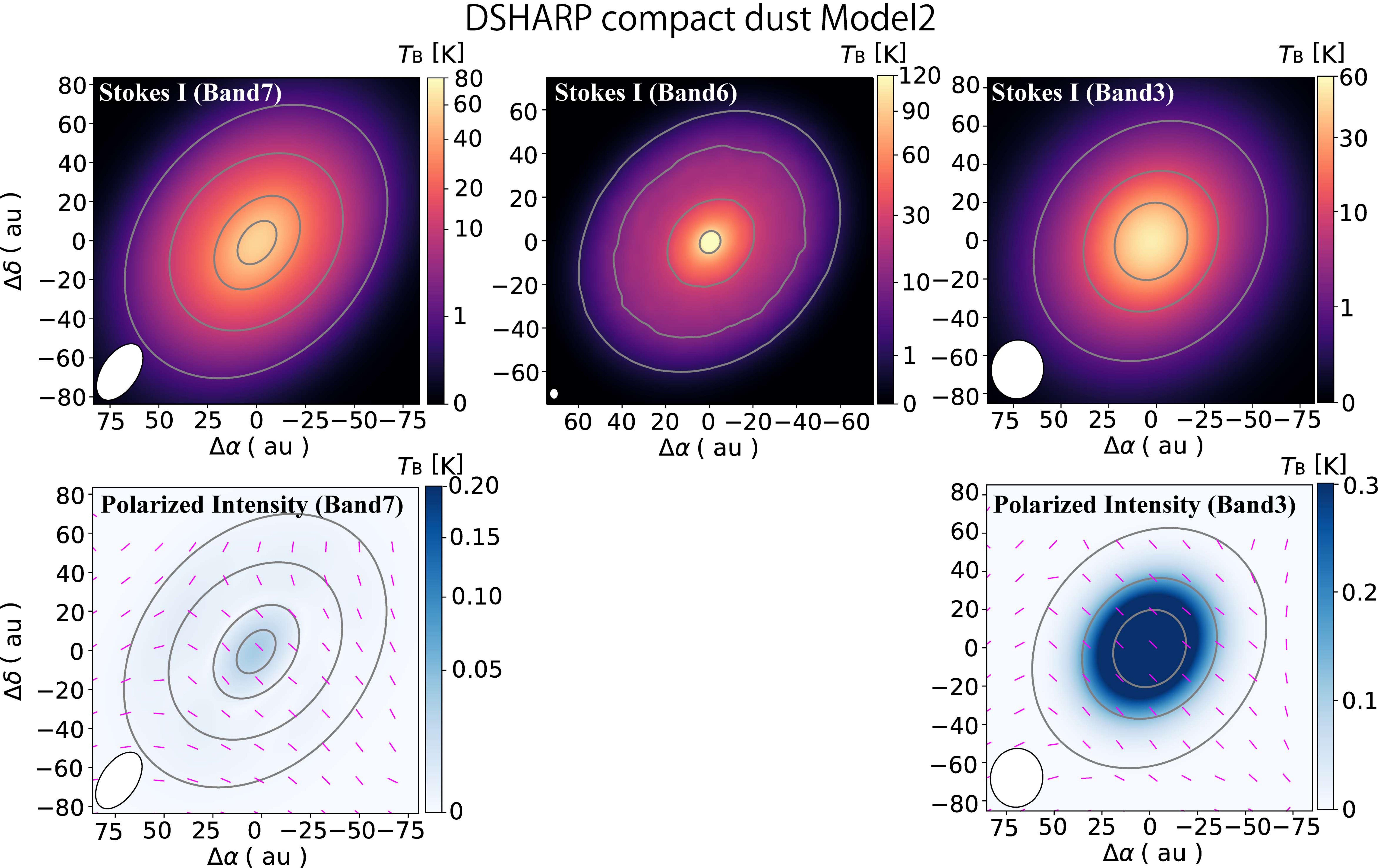}
\end{center}
\caption{Same as Figure \ref{model1_image} but for the model described in Section \ref{sec:model2}.
}
\label{model2_image}
\end{figure*}

Figure \ref{model2_image} shows total intensity and polarization images from the radiative transfer calculations of the model.
Although the total intensity of the model is similar to the observations, the polarized intensity of the model is lower than the observations at 0.87 mm and higher at 3.1 mm.

For the 0.87 mm polarization model, the polarization vectors are parallel to the disk minor axis similar to the observations. However, the polarized intensity reaches only $\lesssim0.01$ K in the central region.  This suggests that the self-scattering would be the reasonable mechanism for the observed polarization, but the maximum grain size of $a_{\rm max}=260$ $\mu$m is inappropriate.
\citep{kat15} showed that the maximum grain size needs to be $a_{\rm max}\sim60-200$ $\mu$m to produce polarization at 0.87 mm.

For the 3.1 mm polarization model, the polarized intensity reaches $T_{\rm B}\approx1.2$ K, which is much higher than the observations of $T_{\rm B}\approx0.2$ K.
This is because the grain size of $a_{\rm max}=260$ $\mu$m is efficient to produce the polarization at 3.1 mm.

Based on the polarization models, we suggest that this model cannot explain the observations because the grain size is too large for self-scattering to be effective at 0.87 mm.
Therefore, we suggest that the smaller dust model described in Section \ref{sec:model1} is more likely than this model.

\subsection{Dependence on Dust Model}\label{sec:sed2}

\begin{figure*}[htbp]
\begin{center}
\includegraphics[width=17.cm,bb=0 0 4000 1368]{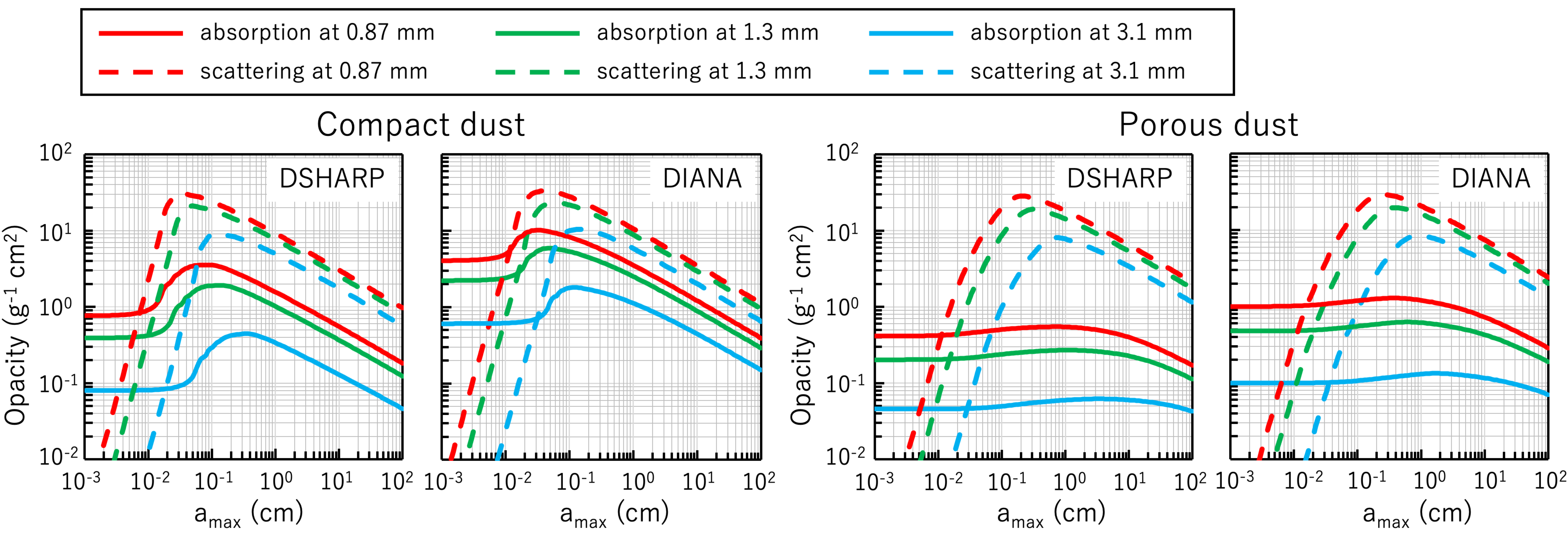}
\end{center}
\caption{absorption (solid lines) and scattering (dashed lines) opacities for the DSHARP and DIANA dust models. The colors indicate the observed wavelengths of 0.87 mm (red), 1.3 mm (green), and 3.1 mm (blue), respectively.
Compact dust means the case without porosity, and porous dust includes 80\% porosity.
}
\label{opacity}
\end{figure*}

\begin{figure*}[htbp]
\begin{center}
\includegraphics[width=17.cm,bb=0 0 1786 2183]{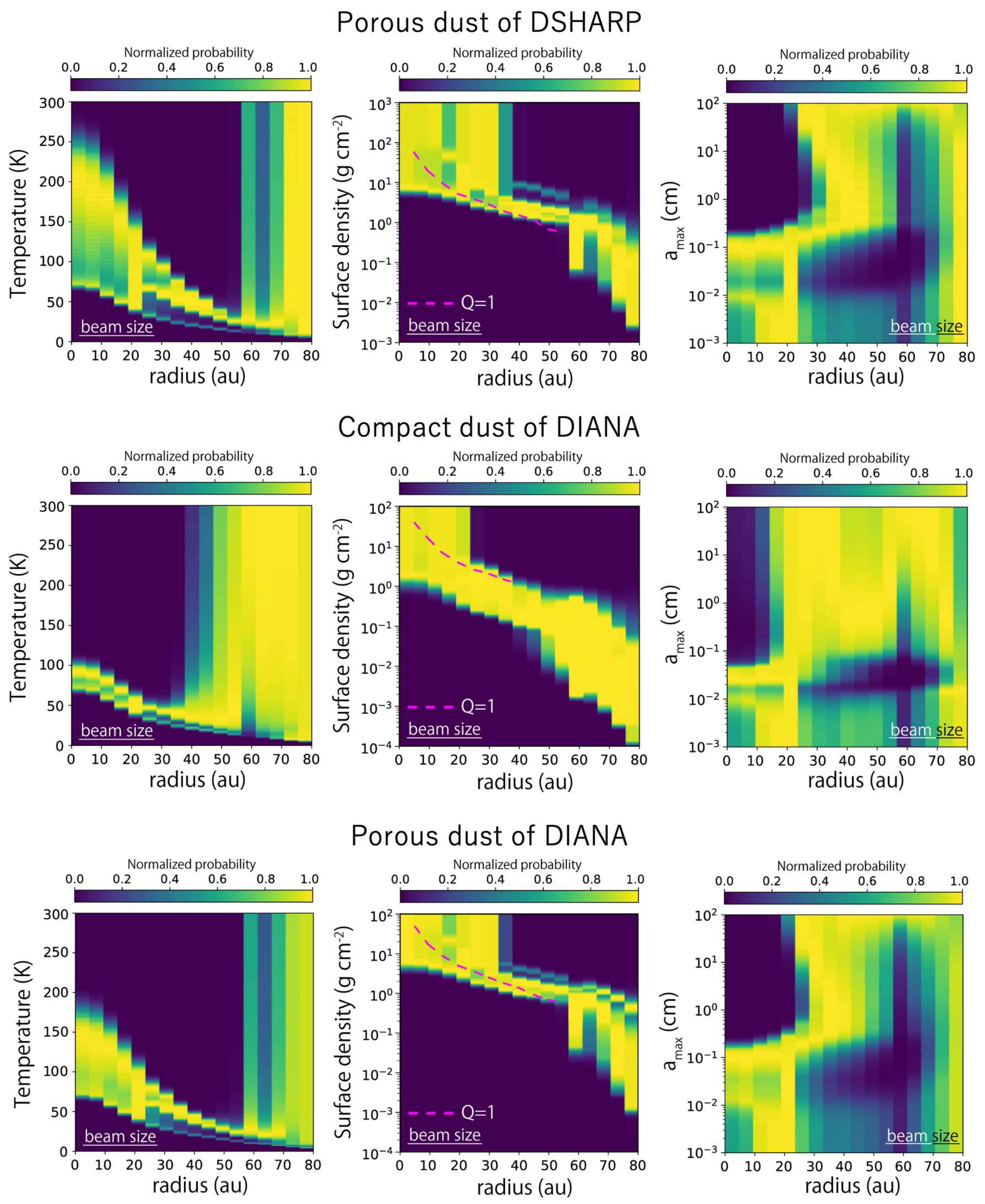}
\end{center}
\caption{Same  as Figure \ref{probability} but for the models of the DSHARP porous dust (80\% porosity) in the upper panel, the DIANA compact dust (without porosity) in the middle panel, and the DIANA porous dust (80\% porosity) in the bottom panel.
}
\label{dustmodelchange}
\end{figure*}

For the previous analysis, we adopted the assumption of spherical dust grains in calculating the dust opacities using the DSHARP model \citep{bir18}. However, the multi-band analysis results and the radiative transfer calculations will be affected by the use of dust models.

We examine two distinct dust models, DSHARP \citep{bir18} and DIANA \citep{woi16}, with compact (without porosity) and 80\% porosity for each model. 
 We assume the power-law size distribution of  $q=-3.5$ for all models.
The publicly available code optool \citep{dom21} is used to calculate the dust opacities.
Figure \ref{opacity} shows the absorption (solid lines) and scattering (dashed lines) opacities for these different dust models as a function of $a_{\rm max}$.
The observed wavelengths of 0.87 mm (red), 1.3 mm (green), and 3.1 mm (blue) are represented by the different colors.
As shown in the figure, the DIANA dust model indicates absorption opacity almost an order of magnitude higher than that of DSHARP because of the different dust compositions of the models.
The DSHARP dust model contains water ice ($\approx20\%$ in mass), silicate ($\approx33\%$), troilite ($\approx7\%$), and refractory organics ($\approx40\%$), while the DIANA dust model comprises of pyroxene (70\% Mg) and carbon in a 0.87/0.13 mass ratio, and a water ice mantle in 20\% of the core mass.

 Here, we investigate the influence of different dust models on the multi-band analysis in Section \ref{subsec:sed2}. The radiative transfer calculations of the DIANA porous dust model are performed to investigate whether the model can agree with the observations in the polarization  in Section \ref{sec:diana}.

\subsubsection{The posterior probability distributions of disk parameters with various dust models}\label{subsec:sed2}

We applied the same method of the multi-band analysis described in Section \ref{sec:sed} but with the above dust models.
We vary $T$, $\Sigma_{\rm d}$, and $a_{\rm max}$ in a $1500\times600\times143$ grid, respectively. $T$ is uniformly spaced from 0.2 to 300 K, $\Sigma_{\rm d}$ is varied in logarithmic space from $10^{-3}$ to 10$^3$ g cm$^{-2}$ for the DSHARP porous dust and from $10^{-4}$ to 10$^2$ g cm$^{-3}$ for the DIANA dust and $a_{\rm max}$ is varied in logarithmic space from 10$^{-3}$ to $10^2$ cm.

The posterior probability distributions of the various disk parameters are shown in Figure \ref{dustmodelchange} in the same way as Figure \ref{probability}.
Our results reveal that the possible solutions of $T$, $\Sigma_{\rm d}$, and $a_{\rm max}$ are different among the dust models.
The temperature and surface density distributions of the porous dust models are found to be higher than those of compact dust models due to the higher albedo and lower absorption opacities. In addition, the $a_{\rm max}$ distribution shows larger dust sizes for the porous dust compared to compact dust in the inner region ($r\lesssim20$ au), which may be attributed to the slight shift in the scattering opacity profile towards larger $a_{\rm max}$ values for the porous dust.

According to the self-scattering theory \citep{kat15}, the grain size of $a_{\rm max}\sim100$ $\mu$m is required for the 0.87 mm dust polarization if the compact dust model is assumed. In contrast, as reported by \citet{taz19}, porous dust aggregates (with a porosity of $\sim99\%$) cannot produce polarized thermal emission. When we assume a porosity of 80\% for the DSHARP and DIANA porous dust models, we find that the grain size of $a_{\rm max}\sim1-2$ mm appears to be a feasible solution for the inner 20 au region. This grain size, when combined with 80\% porosity, can generate polarization similar to the compact dust \citep{taz19}. Hence, the porous dust models may also offer a plausible solution for explaining both the dust continuum and polarization data.
In this case, the grain size of the porous dust models becomes a factor of $3-10$ times larger than the compact dust models.

Based on Figures \ref{probability} and \ref{dustmodelchange}, it appears that the overall trends in the distribution of dust surface density and dust size are similar, regardless of whether we assume the DIANA compact dust or the porous dust models.  Both porous dust models indicate a dust surface density that surpasses the critical value of Toomre $Q=1$ at around 40 au, indicating that the disk is likely gravitationally unstable unless the dust-to-gas mass ratio is increased.
Furthermore, even in the case of DIANA compact dust, the dust surface density indicates a Toomre $Q\sim2.5$ when the dust size is approximately 100 $\mu$m as discussed in the next section.
In addition, the dust size is found to increase with radius in all the models.
In the outer region, the dust size is constrained to be $a_{\rm max}\gtrsim1$ mm.
Further analysis with additional observations with better spatial resolutions will contain the dust size in more detail.
It should be noted that our analysis cannot assess which dust model is the best for the disk because the 3 bands are not enough to reveal the opacity distributions.

\subsubsection{Detailed disk model with DIANA porous dust model}\label{sec:diana}

Since porous dust is suggested as an important pathway for the growth of planetesimals \citep{zha23}, we investigate whether the DIANA porous dust model can reproduce the polarization as well as the continuum emission.
To perform the radiative transfer calculations with the DIANA porous dust model, we assume the similar dust conditions as the previous model of Section \ref{sec:model1}: the smaller dust grains and single power-law temperature of the possible solutions.

Based on Figure \ref{dustmodelchange}, we assume the grain size distribution as follows

\begin{equation}
a_{\rm max} = \begin{cases}
100\ {\rm \mu m}, & r \leq 15 \ {\rm au}, \\
500\ {\rm \mu m}, & 15 < r \leq 25 \ {\rm au}, \\
2\ {\rm mm}, & 25 < r \leq 40 \ {\rm au}, \\
3\ {\rm cm}\ {\rm or}\ 10\ {\rm cm}\ {\rm or}\  100\ {\rm cm} & r > 40 \ {\rm au}. \\
\end{cases}
\label{amax}
\end{equation}

The dust temperature is assumed to be the same as Section \ref{sec:model1}
\begin{equation}
T=294 \Big(\frac{r}{1\ \rm au}\Big)^{-0.6}\ \rm K.
\end{equation}

The dust surface density is assumed to have two exponential power laws to reproduce the change of the intensity slope around $r\approx45$ au found in the 1.3 mm emission. The peak surface density is inferred to be $30$ g cm$^{-2}$ derived from the multi-band analysis. The surface density profile is then given by
\begin{equation}
\Sigma_{\rm d} = \begin{cases}
30 \exp\left[-\Big( \frac{r}{r_{\rm c1}} \Big)^{-\gamma_1}\right]\  {\rm g~cm^{-2}}, & r \leq 40 \ {\rm au}, \\\\
\Sigma_2 \exp\left[-\Big( \frac{r}{r_{\rm c2}} \Big)^{-\gamma_2}\right]\  {\rm g~cm^{-2}}, & r > 40 \ {\rm au}.
\end{cases}
\label{eq:ts}
\end{equation}
With these assumptions, we try to fit the observed intensity profiles of the 0.87 mm, 1.3 mm, and 3.1 mm emission by searching for the parameters of $r_{\rm c1}$, $\gamma_{\rm 1}$, $\Sigma_2$, $r_{\rm c2}$, and $\gamma_2$. 
The best fit values of these parameters are listed in Table \ref{tbl:diana}.

\begin{table}[!ht]
\centering
\vspace{0mm}
\caption{best fit parameters of the DIANA porous dust model}
\begin{tabular}{lcl}
\hline
\hline
fitting parameter & value \\
\hline
$r_{\rm c1}$ &  18 au &\\
$\gamma_{\rm 1}$  & 1.8 & \\
$\Sigma_2$ & 1.1 &for $a_{\rm max}=3$ cm \\
                    &1.6  &for $a_{\rm max}=10$ cm\\
                    &  3.7 &for $a_{\rm max}=100$ cm \\
$r_{\rm c2}$ & 53 au &\\
$\gamma_2$ & 4.8 &\\
\hline
\end{tabular}\\
\label{tbl:diana}
\end{table}

\begin{figure*}[htbp]
\begin{center}
\includegraphics[width=17.cm,bb=0 0 3939 982]{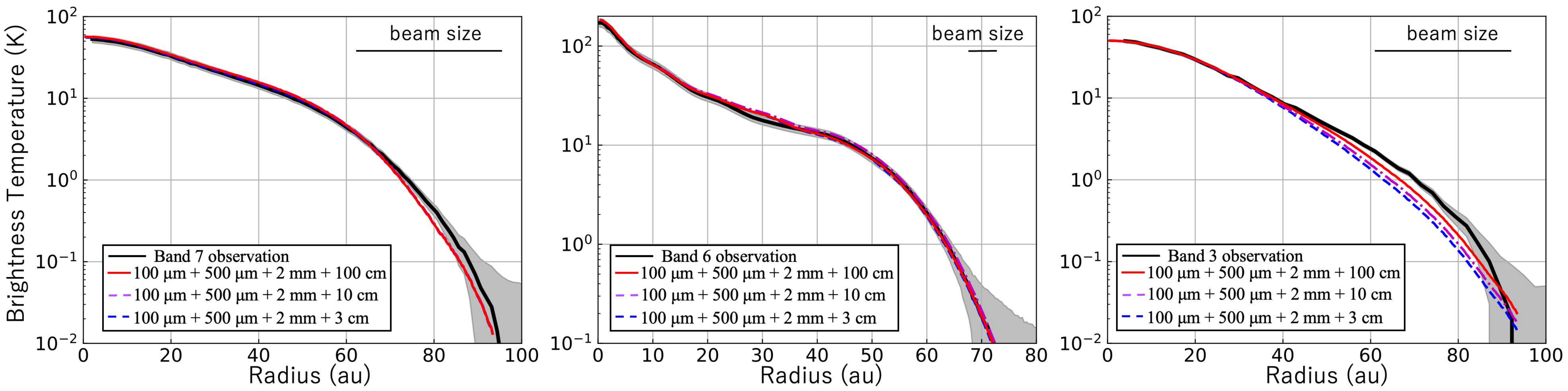}
\end{center}
\caption{Comparison of radial profiles between the observed intensities and model intensities. The model is described in Section \ref{sec:diana}. The black solid lines indicate the observed intensities with uncertainties of the rms noise levels and the flux calibration errors, shown as the shaded region.
The red solid, cyan dashed, and purple dashed lines show the model intensities with dust sizes of 1 cm, 3 mm, and 1 mm in the outer region of the disk ($r\geq40$ au), respectively.
}
\label{diana_profile}
\end{figure*}

Figure \ref{diana_profile} shows the comparisons of the radial profiles of the intensities between the observations and the radiative transfer calculations of our best-fit model. 
We can see that the model reproduces the observed intensities in the inner region ($r\lesssim40$ au) within $3\sigma$ error.
However, these models underpredict the 3.1 mm emission by a factor of $1.5-2$ in the outer region  ($r\gtrsim40$ au), although the larger dust size can better reproduce the observations.
This is because the spectral index is not sensitive to the grain size in the porous dust.
\citet{bir18} showed that the dust opacity index $\beta$ does not decrease with dust size in the porous dust compared to the compact dust.
To better reproduce the 3.1 mm emission, the dust opacity index between 1.3 mm and 3.1 mm needs to be $\beta_{\rm 1.3-3.1\ mm}\sim1.25$, suggesting that the power-law size distribution of $q=-3.5$ needs to be shallower, such as $q=-2.5$ with $a_{\rm max}\sim10$ cm. The shallower power-law is consistent with the drift-limited size distribution rather than the fragmentation-limited one \citep{bir12,bir15}. 
The detailed dust models need to be investigated with further multi-wavelength observations from sub-millimeter to centimeter wavelengths in future.

\begin{figure*}[htbp]
\begin{center}
\includegraphics[width=17.cm,bb=0 0 4746 5936]{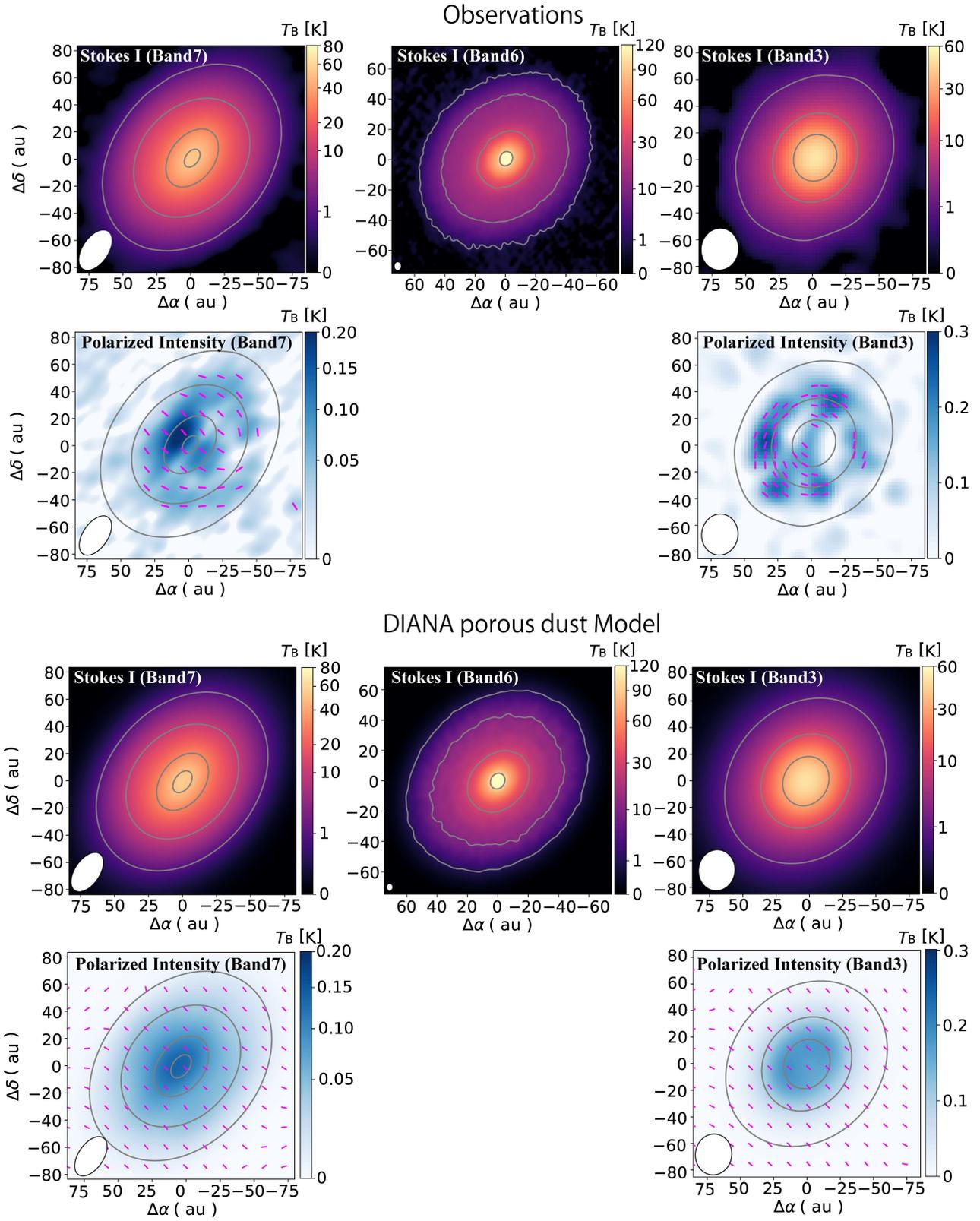}
\end{center}
\caption{Comparison of radial profiles between the observed intensities and model intensities. The black solid lines indicate the observed intensities with uncertainties of the rms noise levels and the flux calibration errors, shown as the shaded region.
The red solid, cyan dashed, and purple dashed lines show the model intensities with dust sizes of 1 cm, 3 mm, and 1 mm in the outer region of the disk ($r\geq40$ au), respectively.
}
\label{diana_image}
\end{figure*}

Figure \ref{diana_image} shows continuum and polarization images of the radiative transfer calculations of our best-fit model. 
The continuum images agree with the observations within $3\sigma$ error in the inner region as described in the radial profile of Figure \ref{diana_profile}.
We found that the polarization images are also similar to the observations.

For the 0.87 mm polarization model, the polarization is widely appeared with a brightness temperature of $T_{\rm B}\sim0.0.5-0.15$ K. This is because porous dust is suggested to have a less
wavelength dependence of the scattering polarization \citep{taz19}.
The model predicts polarization vectors parallel to the disk minor axis, consistent with the observations at $r\lesssim40$ au.
These polarization patterns are consistent with the observations.
However, there are some inconsistencies between the model and observations.
The model shows that the polarization vectors keep parallel to the disk minor axis even in the outer region, while the observed polarization vectors become azimuthal.
This different polarization pattern may be caused by the other polarization mechanisms of radiatively or mechanically aligned non-spherical grains.

The model produces 3.1 mm polarization intensities of $T_{\rm B}\approx 0.2$ K in the inner region of the disk ($r\lesssim40$ au).
The observations also imply a similar brightness temperature in the central position with the polarization vectors parallel to the disk minor axis. 
This suggests that the polarization of the porous dust model is consistent with the observations in the inner region. 
In contrast, our model predicts a brightness temperature lower than 0.01 K for the polarized intensity in the outer region, because the grain size of $a_{\rm max} \gtrsim 3$ cm is too large to produce the polarization for the self-scattering. However, the observations show the brightness temperature as high as $T_{\rm B}\approx0.3$ K with the azimuthal polarization vectors. This indicates that the polarization is caused by other mechanisms rather than the self-scattering.

These comparisons of the DIANA porous dust model and observations suggest that the porous dust can also match the observations not only for the continuum emission (Stokes {\it I}) and but also for the polarization in the inner region  ($r\lesssim40$ au) because the polarization patterns of the model and observations are consistent in the inner region, which are probably caused by the self-scattering.
Even in the outer region, the DIANA porous dust model may be able to match the observations by modifying the shallower power law of the dust size distribution.
Further observations at longer wavelengths will be important to distinguish these compact and porous dust models.

\section{Discussion}\label{sec:dis}

\subsection{Dust enrichment}\label{sec:d2g}

The multi-band analysis showed that the dust surface density is as high as 10 g cm$^{-2}$ or even larger. In addition, our proposed model suggests that the dust surface density in the central region can be $\Sigma_{\rm d}=35$ g cm$^{-2}$ by assuming the DSHARP compact dust.
Such high dust surface density will make the disk gravitationally unstable.

\begin{figure}[htbp]
\begin{center}
\includegraphics[width=8.cm,bb=0 0 1168 1181]{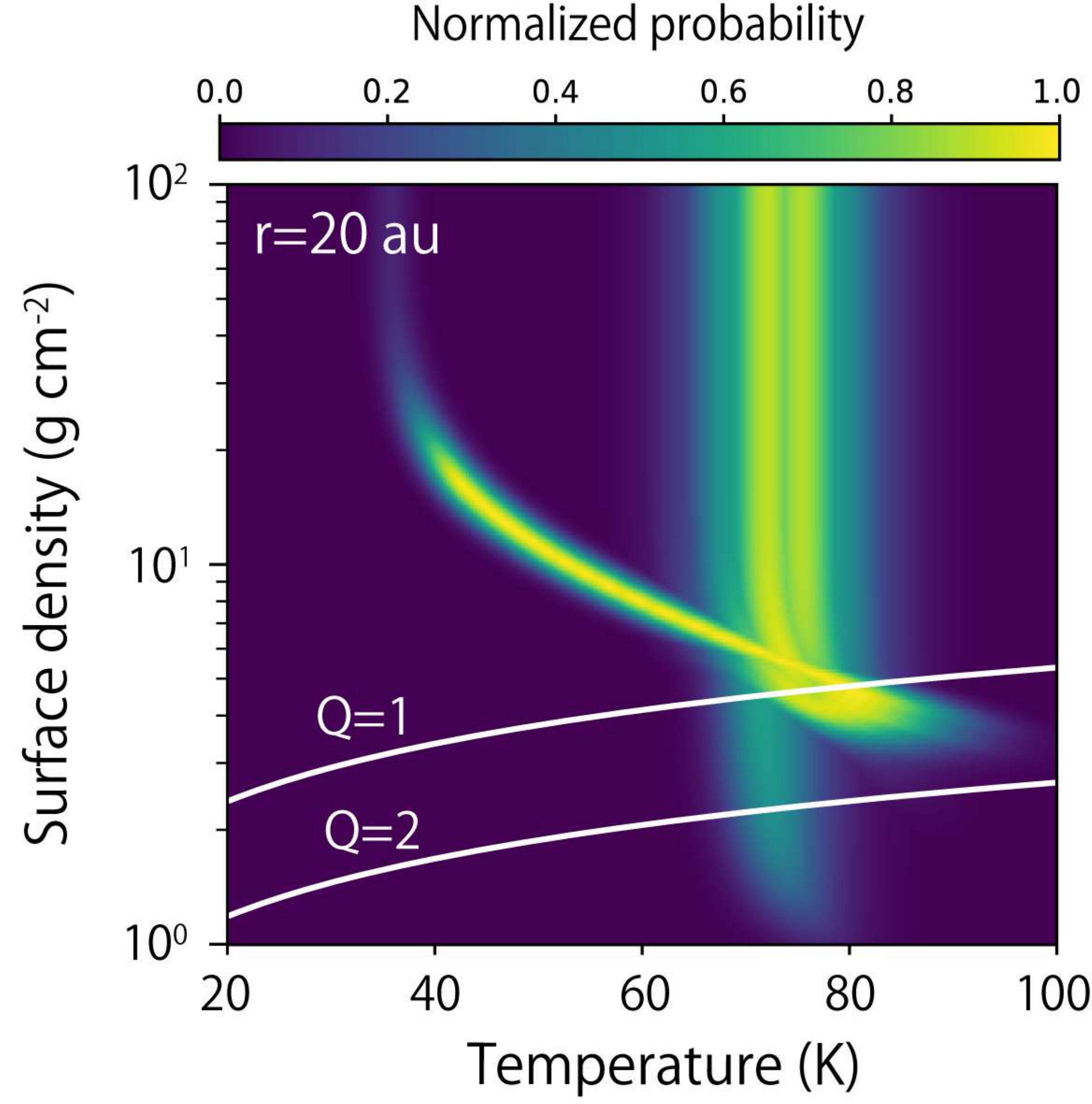}
\end{center}
\caption{Posterior probability distribution of the model parameters fitted to the multi-band observations at a disk radius of 20 au in the DSHARP compact dust.
The parameter space is $\Sigma_{\rm d}$, $T$, and $a_{\rm max}$ but the maximum probabilities in the parameter range of $a_{\rm max} = 10$ $\mu$m $-$ 10 cm are shown.
}
\label{qat20au}
\end{figure}

We compare the observed dust surface density with the critical dust surface density at which the disk becomes gravitationally unstable.
Figure \ref{qat20au} shows the posterior probability distribution of the model parameters of $\Sigma_{\rm d}$ and $T$ at a disk radius of $r=20 $ au in the DSHARP compact dust.
The model parameters also include $a_{\rm max}$, but we plot the maximum value of the probability in the parameter range of  $a_{\rm max}=10$ $\mu$m $-$ $10$ cm.
Figure \ref{qat20au} also plots the critical lines where the Toomre {\it Q} parameter is $1$ and $2$, assuming the dust-to-gas mass ratio of 0.01.

The posterior probability distribution shows that the dust surface density is likely to be higher than the critical dust surface density regardless of the grain size ($Q<2$), indicating that the disk is gravitationally unstable.
However, the 1.3 mm dust continuum image shows no significant substructures such as spirals that would be observed if the disk were gravitationally unstable \citep{bat98,tom17,pan21}.
To keep the disk gravitationally stable and smooth, the dust-to-gas mass ratio needs to be larger.

\begin{figure}[htbp]
\begin{center}
\includegraphics[width=8.cm,bb=0 0 1691 1739]{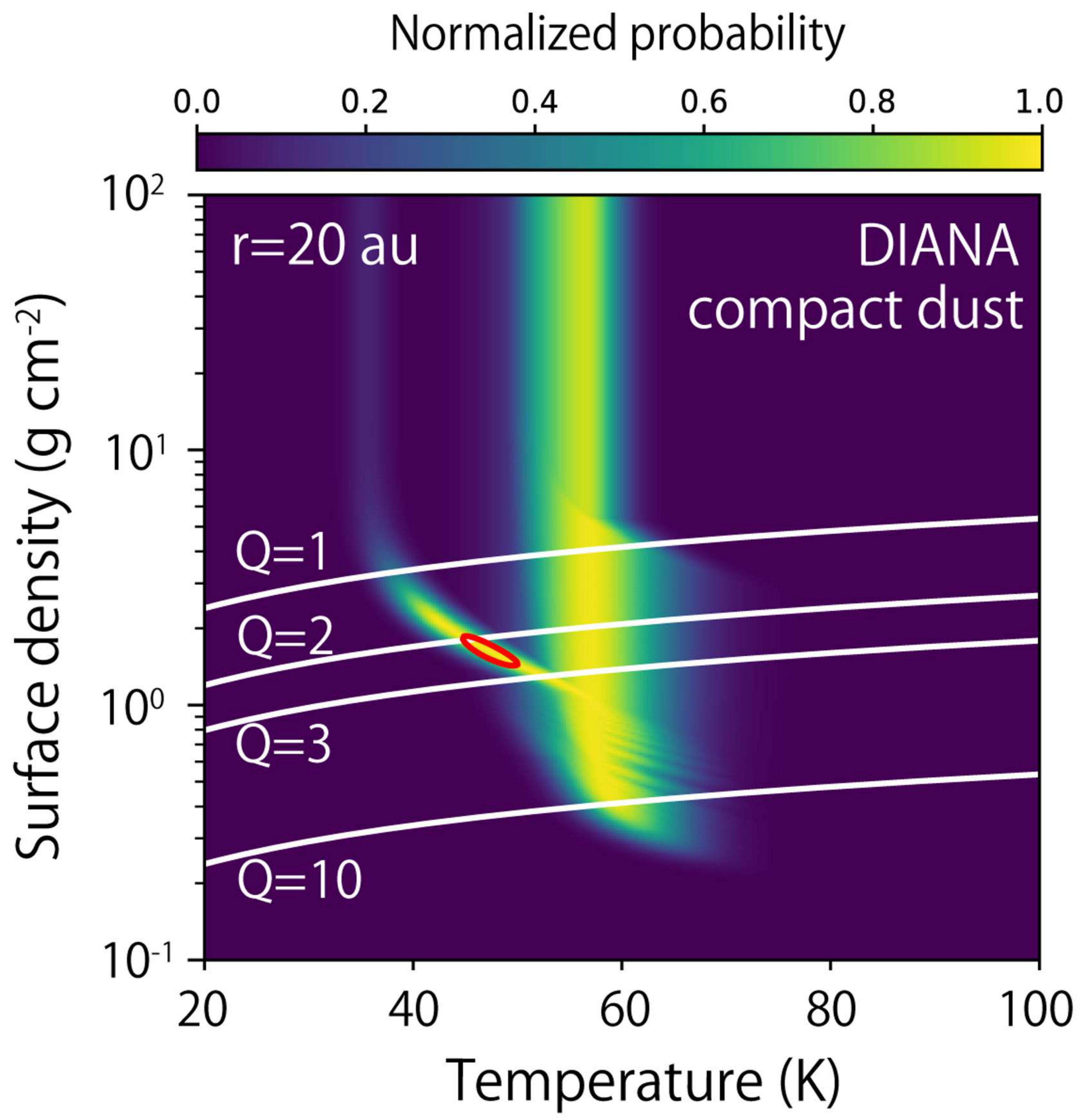}
\end{center}
\caption{Same as Figure \ref{qat20au} but for the DIANA compact dust model.
The red contour indicates the region where the posterior probability is higher than 0.8 with $a_{\rm max}=100$ $\mu$m.
}
\label{qpara_diana}
\end{figure}

\begin{figure}[htbp]
\begin{center}
\includegraphics[width=8.cm,bb=0 0 2593 1907]{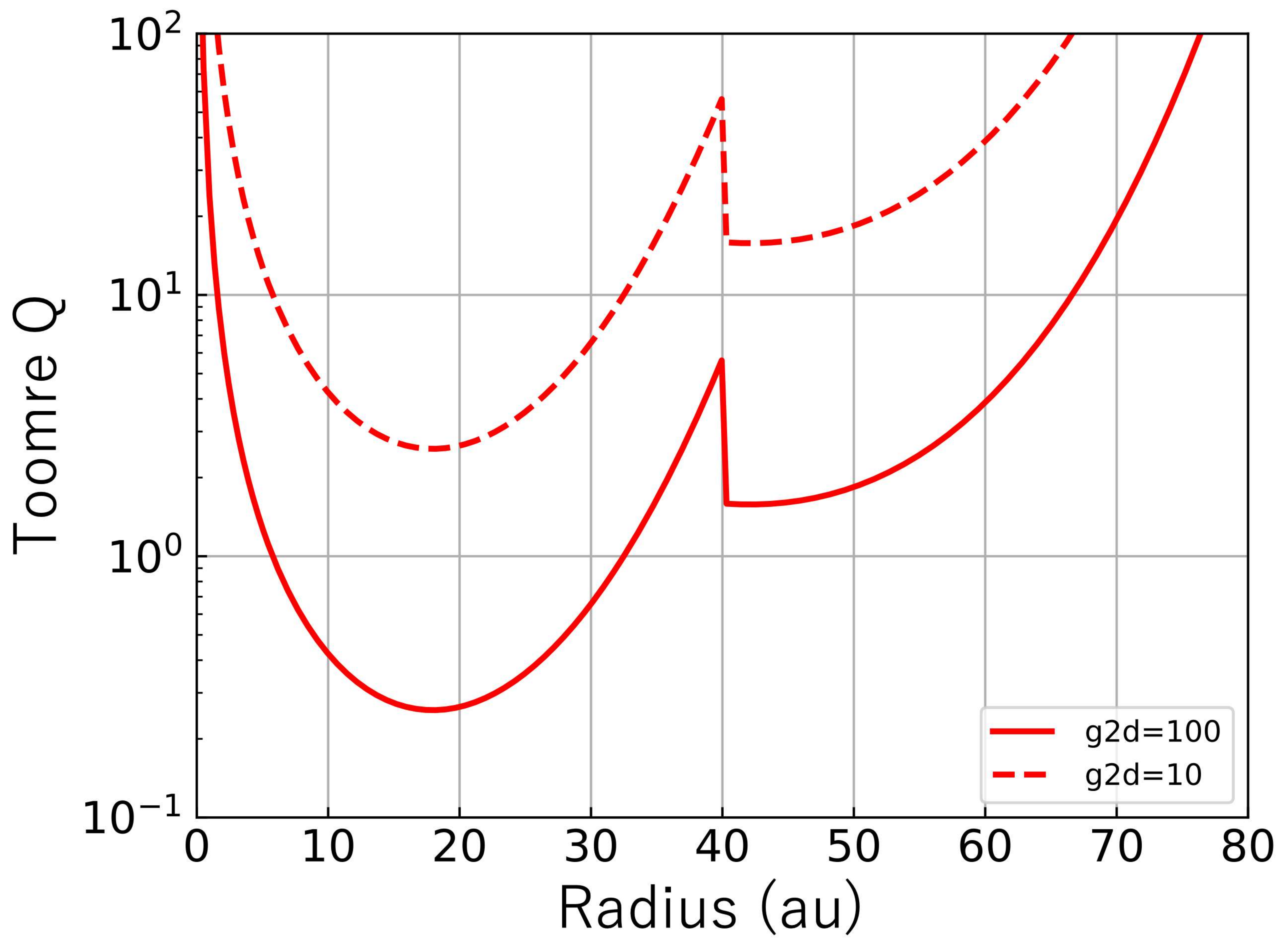}
\end{center}
\caption{Radial profiles of the Toomre {\it Q} values for the model in Section \ref{sec:model1}. The solid line is the case of the dust-to-gas mass ratio of 100, while the dashed line is the dust-to-gas mass ratio of 10.
}
\label{qpara}
\end{figure}

Figure \ref{qpara_diana} shows the same posterior probability distribution as Figure \ref{qat20au} but for the DIANA compact dust model.
The red contour indicates the region where the posterior probability is higher than 0.8 with $a_{\rm max}=100$ $\mu$m.
Since the DIANA compact dust has higher absorption opacities, the surface density is derived to be lower than the DSHARP model.
There are some possible solutions in the dust surface density to keep the disk gravitationally stable ($Q\sim10$).
However, if the dust size is $a_{\rm max}=100$ $\mu$m to generate the polarization caused by self-scattering, the Toomre {\it Q} parameter is found to be $Q\sim2.5$ as shown in the red contour, suggesting that the disk is still marginally unstable if the dust-to-gas mass ratio keeps 0.01.
 Note that the both porous dust models of DSHARP and DIANA indicate that the Toomre {\it Q} parameter is $Q\lesssim1$ around $r\sim20-50$ au because the porous dust models have lower absorption opacities than the compact dust.
Therefore, the dust-to-gas mass ratio needs to be larger in either case of DSHRP or DIANA for the porous dust.

We also investigate the Toomre {\it Q} value of our best-fit model of Section \ref{sec:model1}.
Figure \ref{qpara} shows the radial distribution of the Toomre {\it Q} value of our model.
The solid and dashed lines show the case of the dust-to-gas mass ratio of 0.01 and 0.1, respectively.
We found that the Toomre {\it Q} value becomes as low as 0.2 when the dust-to-gas mass ratio is 0.01 around a 20 au radius.
In other words, the dust-to-gas mass ratio needs to be larger than 0.1 at $r\approx20$ au for $Q\gtrsim2$ to be stable.

The total gas mass of the DG Tau disk is estimated to be $\sim0.015-0.1$ $M_{\odot}$ by the water line emission observed by {\it Herschel}/HIFI observations and with thermos-chemical models \citep{pod13}. From our best-fit model, the total dust mass is derived to be $\sim0.01$ $M_{\odot}$. 
These mass measurements also support the dust enrichment with a dust-to-gas mass ratio of $\gtrsim0.1$.

The dust-to-gas mass ratio of 0.1 indicates that the dust grains are accumulated in the inner region of the disk.
Such dust enrichment may be a common feature of disk evolution because a recent survey study of ALMA multi-band analysis suggests that dust masses have been underestimated by the optically thin assumption \citep{xin23}. 
 The increase of the dust-to-gas mass ratio may be consistent with the recent dust-gas two-fluid non-ideal magnetohydrodynamics (MHD) simulations performed by \citet{tsu23}.
They proposed that the “ash-fall phenomenon” found by \citet{tsu21} will play important role for the evolution of the dust-to-gas mass ratio.

\subsection{Dust Scale height}\label{sec:scale_height}

The 1.3 mm dust continuum image allows us to study the disk morphology thanks to its high spatial resolution. 
We found the symmetric intensity distributions along the disk minor axes in Figure \ref{near_far}.
If a disk is flared, the disk is expected to have an asymmetric intensity along the disk minor axes between the near and far sides \citep{vil20,doi21,oha22}.

\begin{figure*}[htbp]
\begin{center}
\includegraphics[width=17.cm,bb=0 0 3701 2195]{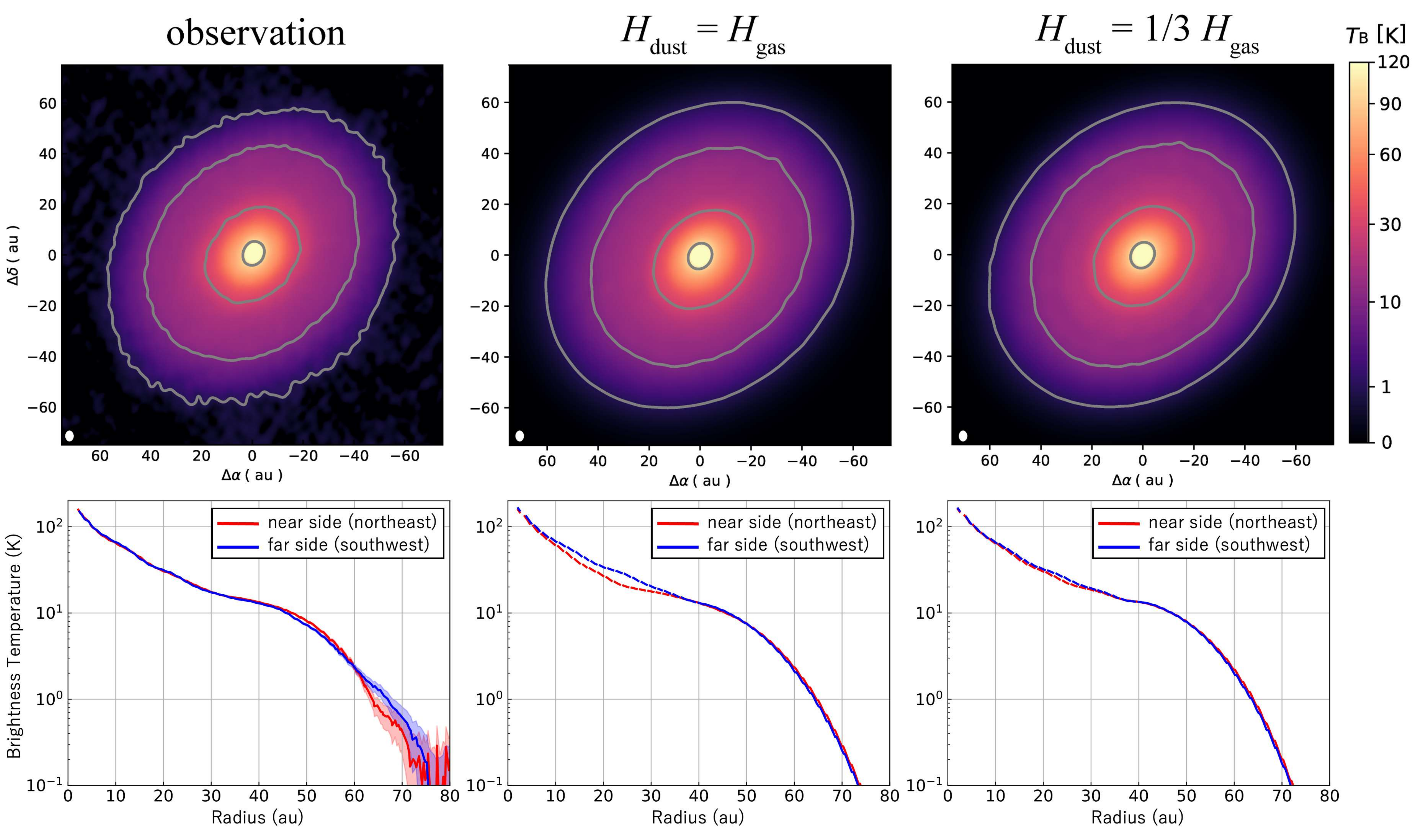}
\end{center}
\caption{1.3 mm dust continuum images (top panel) and radial profiles (bottom panel) of the observation and the model in Section \ref{sec:model1}. The model assumes that the dust scale height is the same with the gas scale height (middle) and is one-third of the gas scale height (right).
The red lines show the radial profiles along the disk minor axis toward the near side, while the blue lines show the radial profiles along the disk minor axis toward the far side.
}
\label{scale_height}
\end{figure*}

Figure \ref{scale_height} shows comparisons of the ALMA 1.3 mm continuum observations and our proposed model. The model considers two cases of a flared disk and a thin disk.
For the flared disk case, the dust scale height ($H_{\rm d}$) is equal to the gas scale height ($H_{\rm g}$). The gas scale height is described in Equation (\ref{eq:scale_height}).
For the thin disk case, the dust scale height is assumed to be one-third of the gas scale height ($H_{\rm d}=1/3\ H_{\rm g}$).
The dust scale heights for the flared and thin disks are estimated to be 2.4 au and 0.8 au, respectively, at a radius of 30 au, corresponding to aspect ratios ($H_{\rm d}/r$) of 0.08 and 0.03.
As shown in Figure \ref{scale_height}, the flared disk has an asymmetric structure along the disk minor axis within $r\leq40$ au.
The near side of the disk is slightly fainter than the far side because the inner region of the near side is obscured by the outer region.
In contrast, the thin disk shows an almost symmetric structure along the disk minor axis similar to the observations.
Therefore, the dust grains of the DG Tau disk should settle to the disk midplane with $H_{\rm d}\lesssim1/3\ H_{\rm g}$.

It should be noted that the dust scale height cannot be constrained outside of a radius of 40 au because neither flared nor thin disks show any difference from the observations.  Both cases of the disk flaring or settling down into the midplane in the outer part of the disk are possible to explain the observations.

The dust scale height is determined by the balance between dust settling and gas turbulence.
By assuming a balance between vertical settling and turbulent diffusion, the dust scale height is written as \citep[e.g.,][]{dub95,you07}
\begin{equation}
        H_{\rm d}=\Big(1+\frac{\rm St}{\alpha}\frac{1+2 \rm St}{1+\rm St}\Big)^{-1/2} H_{\rm g},
\label{eq_scaleheight}
\end{equation}
where St is the Stokes number, ${\rm St}=(\pi\rho a_{\rm max})/(2\Sigma_{\rm gas})$. 
This equation yields $H_{\rm d} \simeq\sqrt{{\alpha}/{\rm St}} H_{\rm g}$ if we assume that ${\rm St}$ is much less than unity.
Using equation (\ref{eq_scaleheight}), the turbulence strength, $\alpha$, is constrained to be $\alpha\lesssim\ {\rm St}/8$ within $r\lesssim40$ au.
Based on our model, we assume the grain size of $a_{\rm max}=160$ $\mu$m, $\Sigma_d=20$ g cm$^{-2}$, and the dust-to-gas mass ratio of 0.1. Then we derive $\alpha\lesssim3\times10^{-5}$ and ${\rm St}\sim2\times10^{-4}$, respectively.

The thin disk and turbulence strength of $\alpha\lesssim3\times10^{-5}$ of the DG Tau disk are consistent with other protoplanetary disks such as HL Tau \citep{pin16,ued21}.
Direct measurements of dust scale heights in several protoplanetary disks also confirm that dust grains are well settled into disk midplanes \citep{vil20,vil22}.
These results suggest that the protoplanetary disks may be in a weakly turbulent state.

The mass accretion rate ($\dot{M}$) can be estimated as
\begin{equation}
\dot{M}=3\pi\alpha_{\rm vis}\frac{c_s^2}{\Omega_{\rm K}}\Sigma_{g},
\end{equation}
where $\alpha_{\rm vis}$ is the $\alpha$-viscosity, $c_s$ is the sound speed, $\Omega_{\rm K}$ is the Kepler frequency, and $\Sigma_{\rm g}$ is the gas surface density.
From the above equation, the $\alpha$-viscosity can be constrained because the mass accretion rate is estimated to be $\dot{M}\sim10^{-7}$ $M_\odot$ yr$^{-1}$\citep{bec10,agr11}. 
By assuming the disk model described in Section \ref{sec:model1}, the dust-to-gas mass ratio of 0.1, and the mass accretion rate of $10^{-7}$ $M_\odot$ yr$^{-1}$ at a 10 au radius, the $\alpha$-viscosity is derived to be $4\times10^{-3}$.
Therefore, the $\alpha$-viscosity is suggested to be much higher than the turbulence strength.
This means that a mechanism other than turbulence (such as a global magnetic field) would cause angular momentum transport in this disk.

\subsection{The origin of the transition of the intensity slope}\label{sec:origin}

Although the DG Tau disk is found to be a smooth structure without significant gaps or ring substructures, the transition of the intensity slope is found at $r\approx45$ au. Our model implies that the dust size and surface density change at this position.

One possibility for changing the dust size and surface density around 40 au would be the CO snowline. 
The CO snowline temperature was estimated to be  $\sim27$ K at the midplane in TW Hya by \citet{zha17}, which is consistent with the temperature at 40 au in DG Tau ($T\approx30$ K).
Indeed, \citet{pod13} and \citet{pod20} suggested that the CO snowline position is $r\approx30$ au, and a change in the dust properties (e.g., dust grain size and opacity).
Our results are consistent with this scenario, and we propose that the dust grains become larger outside the CO snowline.
\citet{oku19} suggested that the grain growth only proceeds to $a_{\rm max}\sim100$ $\mu$m if the dust grains are covered by CO$_2$ ice because the CO$_2$ ice is poorly sticky. 
This may be the case for the inner region ($r\lesssim40$ au) of the DG Tau disk.
In the outer region ($r\gtrsim40$ au), the dust grains are covered by the CO ice, and molecules such as COMs are formed on the dust surface by CO hydrogenation. If these dust compositions are sticky, the grain growth will proceed over $a_{\rm max}\gtrsim3$ mm (${\rm St}\gtrsim0.01$) by coagulation of the dust grains \citep{oha21}.
To confirm this scenario, the stickiness of the COMs ice needs to be investigated.

The coagulation timescale at a 40 au radius can be estimated to be $t_{\rm grow}\sim7\times10^4$ yr \citep{oha21}, which is crudely comparable to the drift timescale of $t_{\rm drift}\sim 3\times10^5\ \rm (St/0.01)^{-1}$ yr \citep{ada76,wei77} for grains with ${\rm St}\sim 0.01-0.1$.
By taking into account that the age of DG Tau ($\sim1$ Myr) is much longer than the coagulation and radial drift time scales and the molecular line emission extends more than the dust distribution, these results suggest that the large dust grains may keep forming and radially drifting by accretion of small dust in the envelope into the disk.
The dust grains will accumulate around the CO snowline position, resulting in an increase in the dust surface density and a larger dust-to-gas mass ratio.
However, \citet{tsu17} showed that the dust-to-gas mass ratio in a protostellar disk becomes as low as 0.001 if the grains, which are continuously supplied from an envelope with the gas, grow and drift inward in the disk. In the case of such low dust-to-gas mass ratio, the Toomre {\it Q} value becomes $Q<<1$ even in the outer part of the DG Tau disk, which would be unlikely. Further simulations such as dust-gas two-fluid non-ideal MHD simulations \citep{tsu23} may help to understand the evolutions of the dust-to-gas mass ratio and dust size distribution in the disk.

Another possibility for the origin of the small dust grains in the inner region could be an outcome of collisional fragmentation.
In this case, large dust grains also exist in the inner region, but the small dust grains can only be observed due to the high optical depth of the dust continuum emission as shown by \citet{oha20}.
This is consistent with the result that the continuum emission is optically thick even at a 3 mm wavelength.
\citet{ued22} showed that shorter wavelength continuum observations can only tracer the upper layer of the disk where small dust grains are located, while longer wavelength observations can trace larger dust grains in the disk midplane due to differential settling of the dust grains if the dust surface density is as high as 10 g cm$^{-2}$.
Such a high dust surface density is consistent with our observations.
Therefore, if collisional fragmentation can produce a large number of dust grains with $a_{\rm max}\sim100$ $\mu$m, this scenario can explain the observation.

\subsection{The initial conditions of planet formation}

A uniqueness of the DG Tau disk is the smooth morphology of the disk structure in the early stage of the star formation.
Recent high-resolution ALMA observations have revealed a variety of structures in protoplanetary disks such as rings, spirals, and crescents \citep{and18}.
In contrast, the DG Tau disk shows no significant substructures even though the disk mass, dust size, and turbulence strength are similar to others such as HL Tau.
This suggests that the DG Tau disk may still be in the early stages of planet formation.

We found that the dust mass of the DG Tau disk is massive, $\sim0.01M_{\odot}$, and the dust-to-gas mass ratio is increasing (0.1 or even higher), which may be important conditions for the process of the planet formation process.
A possible mechanism for the formation of planetesimals is proposed to be the streaming instability, which is triggered by the interaction of solids and gas in a Keplerian disk \citep[e.g.,][]{you05,youd07,joh07}.
The physical conditions that trigger the streaming instability have been investigated, and are suggested to be weak turbulence strength ($\alpha\lesssim10^{-4}$), high dust-to-gas mass ratio, and large pebbles \citep[e.g.,][]{kra19,zhu21}.
The DG Tau disk satisfies the weak turbulence strength and high dust-to-gas mass ratio. Thus, the streaming instability may be a promising scenario for the formation of planetesimals.
In the outer region of the CO snowline, we found large dust grains ($a_{\rm max}\gtrsim3$ mm), which can be explained by the coagulation of the dust grains in the disk \citep{oha21}.
The outer cold region may be a suitable site for grain growth by coagulation due to the sticky composition of the dust.

It may also be possible that planetesimals are already formed in the inner region but the collisional fragmentation produces a large number of the dust grains with $a_{\rm max}\sim100$ $\mu$m whose thermal emission is detected by millimeter wavelength observations.  The recent coagulation simulation showed that planetesimals are rapidly formed in protoplanetary disks \citep{kob21}.
Further observations with longer wavelengths and better spatial resolution will allow us to study the dust size distributions in the disk midplane in detail.

\section{summary}\label{sec:sum}

We have presented the high-spatial-resolution image of the DG Tau disk with ALMA 1.3 mm dust continuum emission.
The observations achieved a spatial resolution of $\approx0\farcs04$, or $\approx5$ au. 
In addition, we use the ALMA archive data of the 0.87 mm and 3.1 mm dust polarization to further analyze the distributions of dust surface density, temperature, and grain size.
The main results are listed below.

\begin{enumerate}

\item The 1.3 mm dust continuum image revealed a geometrically thin and smooth disk structure without a significant substructure such as a ring, but the transition of the radial intensity slope from the power-law-like profile to the exponential-like cut-off is found around a disk radius of $r\approx40-45$ au.

\item The 0.87 mm and 3.1 mm dust polarization data are obtained from the ALMA archive and both data showed that the polarization is caused by self-scattering in the inner region, while it is consistent with the expectations of the thermal emission from radiatively or mechanically aligned non-spherical grains in the outer region.

\item The SED fitting of the radial intensity profiles of the 0.87 mm, 1.3 mm, and 3.1 mm emission shows that the grain size is less than 400 $\mu$m in the inner region ($r\lesssim20$ au), while they are likely to be larger than 3 mm in the outer region ($r\gtrsim40$ au) by assuming the DSHARP compact dust.
The dust surface density is larger than 7 g cm$^{-2}$ in the central region, suggesting that the dust continuum emission is optically thick even at a 3 mm wavelength.

\item The SED fitting and radiative transfer calculations with the various dust models showed a similar trend that dust size is smaller in the inner region and larger in the outer region. The size of the porous dust can be $3-10$ times larger than the compact dust.
To reproduce the 3.1 mm emission in the outer region of the disk ($r\gtrsim 40$ au) with the DIANA porous dust model, the power-law size distribution of $q=-3.5$ needs to be shallower such as $q=-2.5$ with $a_{\rm max}\sim10$ cm, which is consistent with the drift-limited size distribution rather than the fragmentation-limited one.

\item The Toomre {\it Q} parameter is found to be $<2$ at a disk radius of 20 au, assuming a dust-to-gas mass ratio of 0.01, regardless of the grain size. This means a higher dust-to-gas mass ratio is necessary to stabilize the smooth disk structure.
Our best model implies that the dust-to-gas mass ratio could be larger than 0.1.

\item The comparison of our model and 1.3 mm dust continuum images showed that the dust scale height is lower than at least one-third of the gas scale height, suggesting that the turbulent strength is $\alpha\lesssim3\times10^{-5}$.

\item These distributions of dust enrichment, grain sizes, and weak turbulence strength may have significant implications for the formation of planetesimals through mechanisms such as streaming instability.

\item The dust size distribution of the smaller dust size in the inner region ($r\lesssim40$ au) and the larger dust size in the outer region ($r\gtrsim40$ au) can be explained by the CO snowline effect. Inside the CO snowline, the dust grains are covered by CO$_2$ ice, and the grain growth proceeds only to $a_{\rm max}\sim100$ $\mu$m due to the nonsticky CO$_2$ ice. Outside the CO snowline, the dust grains are covered by CO or COMs ice. If these dust compositions are sticky, the grains may grow to $a_{\rm max}\gtrsim3$ mm without the radial drift problem.
Another possibility would be that the small dust grains are produced by the collisional fragmentation.

\end{enumerate}

We gratefully appreciate the comments from the anonymous referee that significantly improved this article.
We thank Ryosuke Tominaga for fruitful discussions.
This paper makes use of the following ALMA data: ADS/JAO.ALMA\#2015.1.01268.S, \\
ADS/JAO.ALMA\#2015.1.00840.S,\\ ADS/JAO.ALMA\#2017.1.00470.S. ALMA is a partnership of ESO (representing its member states), NSF (USA) and NINS (Japan), together with NRC (Canada), MOST and ASIAA (Taiwan), and KASI (Republic of Korea), in cooperation with the Republic of Chile. The Joint ALMA Observatory is operated by ESO, AUI/NRAO and NAOJ. The National Radio Astronomy Observatory is a facility of the National Science Foundation operated under cooperative agreement by Associated Universities, Inc.

This project is also supported by a Grant-in-Aid from Japan Society for the Promotion of Science (KAKENHI: Nos. JP19K23469, JP20K14533, JP20H00182, JP22H01275, JP23H01227).
T.U. acknowledges the support of the DFG-Grant "Inside: inner regions of protoplanetary disks: simulations and observations" (FL 909/5-1).

Data analysis was in part carried out on common use data analysis computer system at the Astronomy Data Center, ADC, of the National Astronomical Observatory of Japan.

\facilities{ALMA}

\software{CASA \citep[][]{casa22}, RADMC-3D \citep{dul12}
          }




\end{document}